\documentclass[journal]{IEEEtran}
\usepackage{amsmath,amssymb,amsthm}
\usepackage{multirow}

\usepackage{enumitem}
\usepackage{cite}
\usepackage{epsfig}
\usepackage[ruled,vlined,lined,linesnumbered]{algorithm2e}
\usepackage{color}
\usepackage{threeparttable}
\usepackage{pgf}

\ifCLASSINFOpdf
\else
\fi
\newcommand{\flp}[1]{`$\mathcal{#1}$'}
\theoremstyle{definition}
\newtheorem{PROP}{Proposition}
\newtheorem{ASMP}{Assumption}

\newcommand{\AEP}[1]{\pgfmathparse{int(round(#1))}\pgfmathresult}
\pgfkeys{/pgf/number format/.cd,fixed,precision=2}
\newcommand{\LTP}[1]{\pgfmathparse{#1}\pgfmathprintnumber[assume math
  mode=true]{\pgfmathresult}}
\begin{document}
%
\title{A multi-mode area-efficient SCL polar decoder}

\IEEEoverridecommandlockouts

\author{
\IEEEauthorblockN{Chenrong Xiong, Jun Lin,~\IEEEmembership{Student~member,~IEEE}
  and Zhiyuan Yan,~\IEEEmembership{Senior~member,~IEEE}}




}


%


\maketitle

\begin{abstract}
Polar codes are of great interest since they are the first provably
capacity-achieving forward error correction codes. To improve throughput and to
reduce decoding latency of polar decoders, maximum likelihood (ML) decoding
units are used by successive cancellation list (SCL) decoders as well as
successive cancellation (SC) decoders. This paper proposes an approximate ML
(AML) decoding unit for SCL decoders first. In particular, we investigate the
distribution of frozen bits of polar codes designed for both the binary erasure
and additive white Gaussian noise channels, and take advantage of the
distribution to reduce the complexity of the AML decoding unit, improving the
area efficiency of SCL decoders. Furthermore, a multi-mode SCL decoder  with
variable list sizes and parallelism is proposed. If high throughput or small
latency is required, the decoder decodes multiple received codewords in parallel
with a small list size. However, if error performance is of higher priority, the
multi-mode decoder switches to a serial mode with a bigger list size. Therefore,
the multi-mode SCL decoder provides a flexible tradeoff between latency,
throughput and error performance, and adapts to different throughput and latency
requirements at the expense of small overhead. Hardware implementation and
synthesis results show that our polar decoders not only have a better area
efficiency but also easily adapt to different communication channels and applications.
\end{abstract}


%
\IEEEpeerreviewmaketitle

\section{Introduction}
Polar codes \cite{5075875}, a breakthrough in coding theory, have
attracted lots of research interest since they achieve the symmetric capacity of
memoryless channels with both binary input \cite{5075875} and nonbinary input \cite{5351487}. A lot of effort has been made to improve the error performance of polar
codes with short or moderate lengths \cite{7055304, 5934670, 6297420, 6283643,
  6708139, ML_polar}, and to improve the hardware
area efficiency of polar decoders \cite{6858413,7114328, JunPolarList, 6327689,
  6804939, 6823099, 6464502, 6680761, 6475198, 6876199, SBSCL2014SiPS, 6986089}.

Maximum likelihood (ML)
decoding algorithms --- the sphere decoding algorithm \cite{6283643}, stack
sphere decoding algorithm \cite{6708139} and a Viterbi 
algorithm \cite{ML_polar} --- can be used to decode polar codes, but their
complexity can be prohibitively high. Compared with ML decoding
algorithms, the successive cancellation (SC) decoding algorithm \cite{5075875}
has a lower complexity at the cost of sub-optimal performance. To improve the
performance of the SC algorithm, the SC list (SCL) decoding algorithm
\cite{6033904} and the CRC-aided SCL (CA-SCL) algorithm \cite{6297420, 7055304}
were proposed. A key drawback of the SC, SCL and CA-SCL algorithms is their
long decoding latency and low decoding throughput, as these algorithms deal
with only one bit at a time: the SC algorithm makes hard bit decisions only one
bit at a time; in the SCL and CA-SCL algorithms, the path expansion is with respect to only one bit. 

To reduce decoding latency and improve throughput of an SC polar decoder,
several algorithms \cite{ParSC, ParSC3, 6464502, 6804939} were proposed to deal with several bits at a time instead of 
only one bit by using ML decoding units, which calculate symbol-wise channel
transition probabilities and make hard decisions for several bits at a time. Based on the SC algorithm, the parallel SC \cite{ParSC}, hybrid ML-SC \cite{ParSC3}, ML
simplified SC (ML-SSC) \cite{6464502} and fast ML-SSC\cite{6804939} algorithms
were proposed. The basic difference of ML decoding units between these four
algorithms is that hybrid ML-SC \cite{ParSC3} and fast ML-SSC\cite{6804939} take
advantage of the distribution of frozen bits to reduce complexity, but
neither parallel SC \cite{ParSC} nor ML-SSC \cite{6464502} algorithms do so.

ML decoding units in
\cite{ParSC2, SBSCL2014SiPS, SDSCL, reduced_latency, reduced_latency_VLSI, 6986089} are also used to improve
throughput of SCL-based decoders and to reduce decoding latency. Instead of making hard decisions in SC-based algorithms,
an ML decoding unit for SCL-based algorithms calculates symbol-wise channel
transition probabilities and performs path expansion and pruning. None of these SCL-based algorithms takes advantage of the 
distribution of frozen bits to reduce complexity of ML decoding units. Therefore, ML decoding
units in these SCL-based algorithms have high complexities. For example, when
the list size is four and the symbol size is eight, the ML decoding unit
accounts for 27\% of the overall decoder area in \cite{SDSCL}.
In \cite{reduced_latency_VLSI},  when the code length is 1024, the area of an ML
decoding unit takes up as much as 62\% of the overall decoder area.



In this paper, we first propose a low-complexity approximate ML (AML) decoding unit by utilizing the
distribution of frozen bits of polar codes and then propose a multi-mode SCL
(MM-SCL) polar decoder to support variable throughput and latency. Our main contributions are:

\begin{itemize}
\item The divide-and-conquer method in \cite{ParSC3} is applied in the
  probability domain to simplify the ML unit for SC-based algorithms. By
  extending this idea, a divide-and-conquer AML decoding unit for SCL-based
  algorithms is proposed by considering the distribution of frozen bits. Its
  computational complexity is greatly smaller than those of existing ML decoding units for SCL-based algorithms. When an appropriate
  design parameter for the divide-and-conquer AML decoding unit is selected, the
  SCL decoder has negligible performance loss.
\item The distribution of frozen bits of polar codes is analyzed from the viewpoint of code
  construction. We show that there are only a small number of frozen-location
  patterns for polar codes constructed by a method proposed by Ar{\i}kan in
  \cite{4542778} and a method in \cite{6823688}. 
\item Since only a small number of frozen-location patterns exist in polar
  codes, the divide-and-conquer AML decoding unit for SCL-based algorithms is
  simplified further. A low-complexity hardware implementation for the simplified
  divide-and-conquer AML decoding unit, the LC-AML decoding unit, is proposed. Synthesis results show that
  by taking advantage of a small number of frozen-location patterns, our CA-SCL decoder with the
  LC-AML unit has a better area efficiency than existing SCL decoders, while working for all channel conditions.
\item An MM-SCL polar decoder is also proposed. This decoder supports SCL
  algorithms with different list sizes and parallelism. When a high throughput or small
  latency is needed, the MM-SCL decoder decodes multiple
  received codewords in parallel with a small list size. If a good performance is required, the MM-SCL
  decoder switches to a mode with a greater list size to decode only one receive
  codeword. Therefore, the MM-SCL polar decoder provides a flexible tradeoff
  between latency, throughput and performance, and consequently adapts to different throughput
  and latency requirements at the expense of small overhead.
\end{itemize}

Our proposed divide-and-conquer AML decoding unit for SCL-based algorithm is an
extension of the method for SC-based algorithm in \cite{ParSC3}.
However, by investigating the distribution of frozen bits of polar
codes, we 
reduce the complexity of the ML decoding unit further. Existing
ML decoding units for SCL decoders \cite{ParSC2, SBSCL2014SiPS, SDSCL, reduced_latency,
  reduced_latency_VLSI} perform list pruning function after all the symbol-wise 
channel transition probabilities are calculated, whereas the LC-AML decoding
unit sorts intermediate calculation results generated by the recursive channel combination method \cite{SBSCL2014SiPS}, and
consequently reduces the number of symbol-wise channel transition probabilities
dealt with by the list pruning function. Hence, the
proposed LC-AML decoding unit has a much smaller complexity. The performance
degradation due to the proposed LC-AML is the same as that in \cite{SBSCL2014SiPS,
  SDSCL}. Although the ML decoding unit in \cite{ParSC2} has no performance
degradation, its complexity grows quickly as the list size and symbol size
increase. The performance degradation of the ML decoding unit in 
\cite{reduced_latency, reduced_latency_VLSI} depends on its design parameters.
 
Many applications, such as modern wireless or wireline communication system,
require variable data rate transmission and have stringent latency requirements. 
As a potential
candidate of FEC technique for future communication systems, a polar decoder
supporting variable data rate and variable decoding latency is
desired. Unfortunately, existing polar decoders
provide only fixed latency and throughput (data rate).
To the best of our knowledge, the proposed MM-SCL decoder is the \emph{first}
polar decoder with variable throughput and decoding latency given a polar code. 

The rest of this paper is organized as follows. In Section~\ref{sec:intro}, polar
codes as well as construction methods for polar codes and existing
ML decoding units are reviewed. In Section~\ref{sec:LC_MLD_SCL}, first, the
divide-and-conquer method is applied in the probability domain for the ML unit
of SC-based algorithms. Then, the divide-and-conquer AML 
decoding unit for SCL-based algorithms is proposed, and its computational complexity
is also analyzed. In Section~\ref{sec:data_pattern}, frozen-location patterns for polar codes are
investigated.
In Section~\ref{sec:ML_design}, 
a hardware design of the LC-AML unit is proposed, and an area-efficient CA-SCL decoder with the LC-AML unit is implemented as
well. The hardware implementation and synthesis results for the area-efficient SCL decoder
are also presented in this section. In Section~\ref{sec:mm_SCL}, the MM-SCL
decoder, its hardware implementation and synthesis results are
presented. Finally, some conclusions are provided in
Section~\ref{sec:conclusion}.

\section{Preliminaries}
\label{sec:intro}
\subsection{Notations}
Suppose $\mathbf{u}$ represents a binary vector
$(u_1,u_2,\cdots,u_N)$. $u_a^b$ denotes a binary subvector
$(u_a,u_{a+1},\cdots,u_{b-1},u_b)$ of $\mathbf{u}$ for $1 \leq a,b \leq N$; if
$a>b$, $u_a^b$ is regarded as void. $u_{a,o}^{b}$ and $u_{a,e}^{b}$ represent
the subvectors of $u_a^b$ with odd
and even indices, respectively. For an index set $\mathcal{A} \subseteq
\mathcal{I}=\{1,2,\cdots, N\}$, its complement in $\mathcal{I}$ is
denoted as $\mathcal{A}^c$. The subvector of $\mathbf{u}=u_1^N$ restricted to
$\mathcal{A}$ is represented by $\mathbf{u}_{\mathcal{A}}=(u_i: 0 < i \leq N, i\in
\mathcal{A})$. 

\subsection{Polar codes}

For an $(N,K)$ polar code, the code length $N$ is a power of two, i.e., $N=2^n$ for $n
>0$. The data bit sequence, represented by $\mathbf{u}=u_1^N$, is
divided into two parts: a $K$-element part $\mathbf{u}_{\mathcal{A}}$ which
carries information bits, and $\mathbf{u}_{\mathcal{A}^c}$ whose elements
(called frozen bits) are set to zero. The
corresponding encoded bit sequence $\mathbf{x}=x_1^N$ is generated by $\mathbf{x} = \mathbf{u}B_NF^{\otimes n}$,
where
 $B_N$ is the $N\times N$ bit-reversal permutation matrix, $F=\left[
\begin{smallmatrix}
1 & 0 \\
1 & 1
\end{smallmatrix}
\right]$, and $F^{\otimes
  n}$ is the $n$-th Kronecker power of $F$ \cite{5075875}.

\subsection{Construction Methods of Polar Codes}
An essential problem for constructing a polar code is to determine the locations of
frozen bits (elements of $\mathcal{A}^c$). For the BEC with an erasure probability $\epsilon$ $(0 < \epsilon
< 1)$, assuming $z_{0,1} =
\epsilon$, the following recursions \cite{4542778} are used to construct an $(N,
K)$ polar code, where $N=2^n$ and $ 0<K<N$: 
\begin{gather}
\label{eq:BEC_cons}
z_{i,2j-1} = 2z_{i-1,j} - z^2_{i-1,j}, \\
\label{eq:BEC_cons2}
z_{i,2j} = z^2_{i-1,j},
\end{gather}
where $1 \leq i \leq n$. Then $\mathcal{A}^c$ is chosen such that $\sum_{j\in\mathcal{A}^c}z_{n,j}$
is maximal and $\lvert \mathcal{A}^c \rvert = N-K$.

For the AWGN channel and a given initial value of $z_{0,1}$, which is determined
by a desired signal-to-noise ratio (SNR), the following recursive method \cite{6823688} based on
Gaussian approximation is used for $1 \leq i \leq n$:
\begin{gather}
\label{eq:AWGN_construction}
z_{i,2j-1} = \tau^{-1}\Bigl(1-\bigl(1-\tau(z_{i-1,j})\bigr)^2\Bigr), \\
\label{eq:AWGN_construction2}
z_{i,2j} = 2z_{i-1,j},
\end{gather}
where 
\begin{equation}
\label{eq:phi_func}
\tau(x) = \left\{
  \begin{array}{l l}
    e^{-0.4527x^{0.86}+0.0218}, & \quad 0<x<10,\\
    \sqrt{\frac{\pi}{x}}e^{-\frac{x}{4}}(1-\frac{10}{7x}) & \quad x \geq 10.
  \end{array} \right.
\end{equation}
In this case, $\mathcal{A}^c$ is chosen such that $\sum_{j\in\mathcal{A}^c}z_{n,j}$
is minimal and $\lvert \mathcal{A}^c \rvert = N-K$.

\subsection{Existing ML Decoding Units for Polar Decoders}
When $\mathbf{x}= \mathbf{u}B_NF^{\otimes n}$ is transmitted, suppose the received word is
$\mathbf{y}=y_1^N$ and the symbol size is $M=2^m$. A symbol-decision
\cite{SBSCL2014SiPS} ML
decoding unit first calculates symbol-wise channel transition probabilities, 
${\Pr}(\mathbf{y},\hat{u}_1^{jM}|u_{jM+1}^{jM+M})$ $(0\leq j <
\frac{N}{M})$, then makes a symbol-wise ML
decision for SC-based decoders or chooses the $L$ most reliable paths for SCL-based
decoders. Here, $\hat{u}_1^{jM}$ is the previously estimated bits.

There are three methods to calculate the symbol-wise channel transition probabilities, and all of
them do not take advantage of the distribution of frozen bits. The first
\cite{ParSC, ParSC2, 6464502, reduced_latency} is based on an $M$-element product of
bit-wise channel transition probabilities, called directing mapping method (DMM): 
\begin{equation}
\label{eq:DCS}
\begin{split}
{\Pr}&(\mathbf{y},\hat{u}_1^{iM-M}|u_{iM-M+1}^{iM}) = \\
&\prod_{j=0}^{M-1}{\Pr}(y_{j\frac{N}{M}+1}^{(j+1)\frac{N}{M}},\hat{w}_{1+j\frac{N}{M}}^{(i-1)+j\frac{N}{M}}|w_{i+j\frac{N}{M}}),
\end{split}
\end{equation}
where $u_{iM-M+1}^{iM}=(w_i,w_{i+\frac{N}{M}},\cdots,w_{i+N-\frac{N}{M}})B_MF^{\otimes
  m}$ for $1 \leq i \leq \frac{N}{M}$, and
$\hat{w}_{1+j\frac{N}{M}}^{(i-1)+j\frac{N}{M}}$ is the previously estimated bit
vector of ${w}_{1+j\frac{N}{M}}^{(i-1)+j\frac{N}{M}}$.

The second \cite{SBSCL2014SiPS}, called as the recursive channel
combination (RCC) method, is based on a product of
symbol-wise channel transition probabilities recursively,
\begin{equation}
\begin{split}
{\Pr}(y_{1}^{2\Lambda},\hat{u}_1^{i\Phi}|u_{i\Phi+1}^{i\Phi+\Phi})=&{\Pr}(y_1^{\Lambda},\hat{u}_{1,o}^{i\Phi}\oplus
\hat{u}_{1,e}^{i\Phi}|u_{i\Phi+1,o}^{i\Phi+\Phi}\oplus
u_{i\Phi+1,e}^{i\Phi+\Phi})\\
&\cdot{\Pr}(y_{\Lambda+1}^{2\Lambda},\hat{u}_{1,e}^{i\Phi}|u_{i\Phi+1,e}^{i\Phi+\Phi}),
\end{split}
\end{equation}
where $1 \leq \phi \leq m$, $0 \leq \lambda <n$, $\Lambda=2^{\lambda}$, and $\Phi=2^{\phi}$.

The third \cite{reduced_latency_VLSI} is a hybrid method by applying the DMM
first and then the RCC method, referred to as the DRH method.

Based on the distribution of frozen bits, some data symbols
in \cite{6804939} are considered as some special constituent codes, such as repetition
codes and single-parity-check nodes. Different methods were proposed to
deal with different constituent codes.

Furthermore, an ML decoding unit in \cite{ParSC3} with the
divide-and-conquer method was proposed for SC algorithms based on an empirical
assumption \cite{ParSC3}:

\ASMP{For a well designed polar code, there is no such case that $u_{2i-1}$ is an
information bit and $u_{2i}$ is a frozen bit, for any $1\leq i \leq
\frac{N}{2}$.}

Based on this assumption and the divide-and-conquer method, a simplified ML unit
was proposed in \cite{ParSC3}. Moreover, a recursive way of the divide-and-conquer method was
proposed in \cite{ParSC3}, but it is not suitable for hardware implementation
since it is for a large symbol size, which has a very high complexity
for hardware implementation.

How to take advantage of frozen-location patterns to reduce complexity of ML decoding
units has been discussed in \cite{6804939} and \cite{ParSC3} for SC-based
algorithms, but it has not been investigated yet for SCL-based algorithms.


\section{Divide-and-Conquer AML Decoding Unit}
\label{sec:LC_MLD_SCL}
The simplified ML unit in \cite{ParSC3} is based on the Euclidean distance since
an AWGN channel is assumed. Here, we first apply the divide-and-conquer method
in the probability domain and reformulate the ML unit for SC-based
algorithms. This simplified ML unit in the probability domain is slightly more
general than that in \cite{ParSC3}, because it is applicable to both AWGN channels
and other channels. We then extend the simplified ML unit in the probability domain to
SCL-based algorithms. 

\subsection{Reformulation of the divide-and-conquer ML unit for SC-based
  algorithms \cite{ParSC3} in the probability domain}
For the ease of discussion, a string vector 
$\mathbb{S}_a^b=$`$\mathcal{S}_{a}\cdots\mathcal{S}_{b}$' (for $1 \leq a \leq b
\leq N$) is introduced to represent a frozen-location pattern of symbol $u_a^b$. If $u_{j}$ $(a \leq j\leq b)$ is an information bit, $S_{j}$ is
denoted as `$\mathcal{D}$'. Otherwise, $S_{j}$ as `$\mathcal{F}$'. Consider a
toy example of a four-bit symbol $u_{1}^{4}$. Assuming $u_{1}$ and
$u_{3}$ are frozen bits, and $u_{2}$ and $u_{4}$ are information
bits. Then the frozen-location pattern $\mathbb{S}_1^4$ of $u_{1}^{4}$ is
`$\mathcal{FDFD}$'. Obviously, for all $M$-bit symbols, there are $2^M$ frozen-location patterns. 

Based on Assumption 1, $u_{jM+1}^{jM+M} (0\leq j < \frac{N}{M})$ can be divided into
$\frac{M}{2}$ pairs, $u_{jM+2i-1}$ and $u_{jM+2i}$ for $1 \leq i \leq
\frac{M}{2}$. In theory, any pair of $u_{jM+2i-1}$ and $u_{jM+2i}$ have four
possibilities. \flp{FF} is trivial. Under Assumption
1, \flp{DF} is not possible. Hence, in \cite{ParSC3}, only two remaining possibilities
are considered: \flp{FD} and \flp{DD}. Let $\Omega_{01}^{(j)}$ represent the
index set of $i$ that $u_{jM+2i-1}^{jM+2i}$ is \flp{FD}. $\Omega_{11}^{(j)}$ represents the index set of
$i$ that $u_{jM+2i-1}^{jM+2i}$ is \flp{DD}.


For SC-based algorithms, the maximum of $2^{\lvert\Omega_{01}^{(j)}\rvert + 2\lvert\Omega_{11}^{(j)}\rvert}$ values of
$T(u_{jM+1}^{jM+M})\triangleq$ $\Pr(\mathbf{y},\hat{u}_1^{jM}|u_{jM+1}^{jM+M})$
needs to be found. Based on the RCC method \cite{SBSCL2014SiPS},
$T(u_{jM+1}^{jM+M})=T_1(v_{\frac{jM}{2}+1}^{\frac{jM+M}{2}})\times
T_2(u_{jM+1,e}^{jM+M})$, where 
$v_1^{\frac{N}{2}}$$\triangleq$$u_{1,o}^{N}\oplus u_{1,e}^{N}$, $T_1(v_{\frac{jM}{2}+1}^{\frac{jM+M}{2}})\triangleq\Pr(y_1^{\frac{N}{2}},\hat{v}_{1}^{\frac{jM}{2}}|v_{\frac{jM}{2}+1}^{\frac{jM+M}{2}})$,
and
$T_2(u_{jM+1,e}^{jM+M})$$\triangleq$$\Pr(y_{\frac{N}{2}+1}^{N},\hat{u}_{1,e}^{jM}|u_{jM+1,e}^{jM+M})$. 
The possible values of $T(u_{jM+1}^{jM+M})$ can be divided into $2^{\lvert\Omega_{01}^{(j)}\rvert}$ groups
based on $\Omega_{01}^{(j)}$, each with $2^{2\lvert\Omega_{11}^{(j)}\rvert}$
values. In each group, for $i\in\Omega_{11}^{(j)}$, since $v_{\frac{jM}{2}+i}$ and
$u_{jM+2i}$ are independent, $\max(T)=\max(T_1)\max(T_2)$. Then, the
maximum of $2^{\lvert\Omega_{01}^{(j)}\rvert}$ values generated in the previous step is found.
Therefore, if $\Omega_{01}^{(j)}=\varnothing$,

\begin{equation}
\label{eq:MLD_SC_nothing}
\underset{u_{jM+1}^{jM+M}}\max{(T)}=\underset{\begin{subarray}{c}u_{jM+2i-1}^{jM+2i}\\i\in
  \Omega_{11}^{(j)}\end{subarray}}\max{(T_1)} \times\underset{\begin{subarray}{c}u_{jM+2i}\\i\in \Omega_{11}^{(j)}\end{subarray}}\max{(T_2)};
\end{equation}

\noindent otherwise, 
\begin{equation}
\label{eq:MLD_SC}
\underset{u_{jM+1}^{jM+M}}\max{(T)}= \underset{\begin{subarray}{c}u_{jM+2i}\\i\in \Omega_{01}^{(j)}\end{subarray}}\max\left(\underset{\begin{subarray}{c}u_{jM+2i-1}^{jM+2i}\\i\in
  \Omega_{11}^{(j)}\end{subarray}}\max{(T_1)}\times \underset{\begin{subarray}{c}u_{jM+2i}\\i\in \Omega_{11}^{(j)}\end{subarray}}\max{(T_2)}\right).
\end{equation}



\begin{figure*}[htbp]
\centering
\includegraphics[width=18cm]{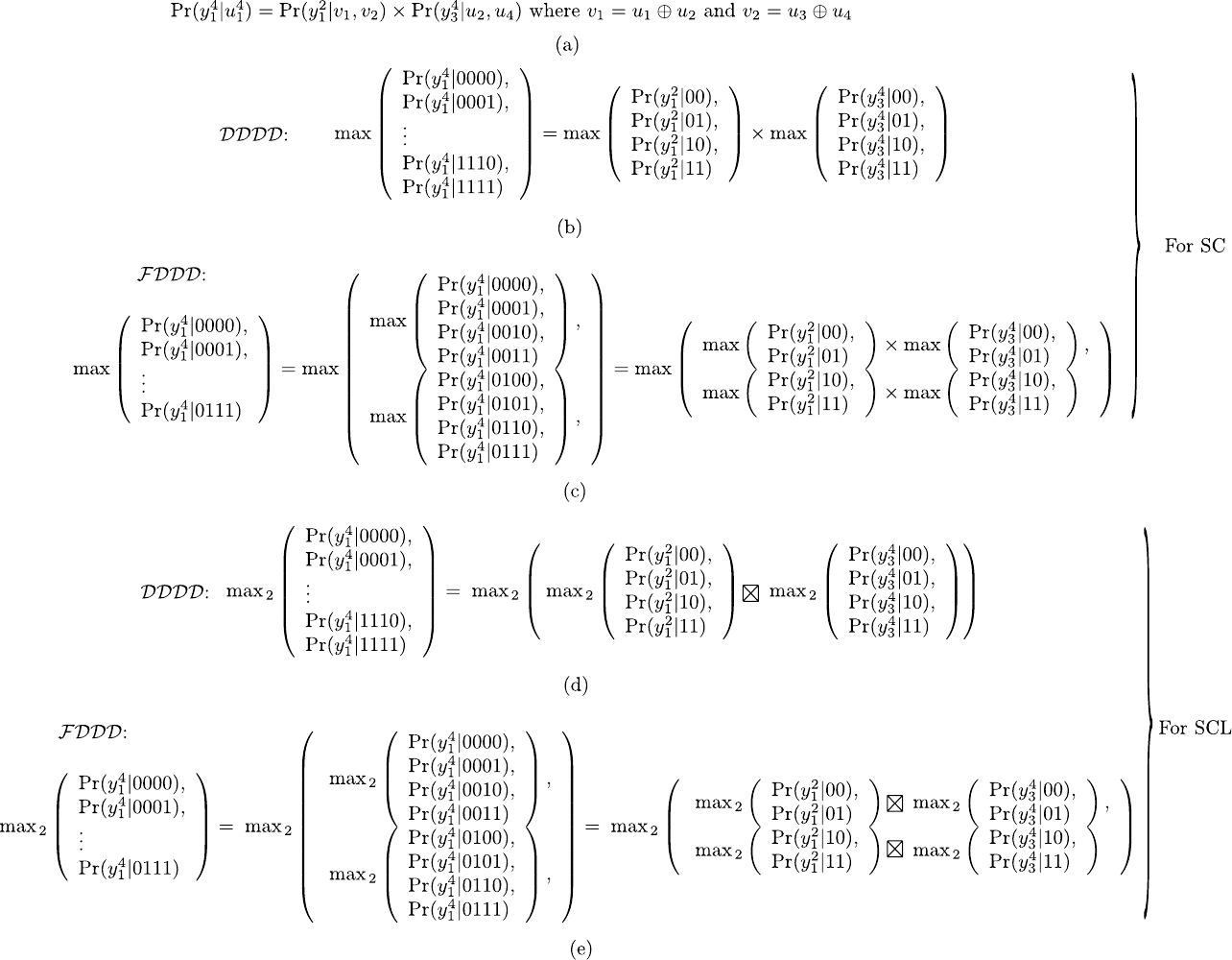}
\caption{Examples for calculating $\max()$ and $\rm{max}_2()$ functions when $M=4$,
  $N=4$, $q=2$, and $j=0$, (a) the calculation rule for $\Pr(y_1^4|u_1^4)$, (b) the
  calculation of $\max(\Pr(y_1^4|u_1^4))$ in \cite{ParSC3} when the frozen-location pattern is `$\mathcal{DDDD}$'
  , (c) the calculation of $\max(\Pr(y_1^4|u_1^4))$ in \cite{ParSC3} when the
  frozen-location pattern is `$\mathcal{FDDD}$', (d) the
  calculation of $\rm{max}_2(\Pr(y_1^4|u_1^4))$ when the frozen-location pattern is `$\mathcal{DDDD}$'
  , and (e) the
  calculation of $\rm{max}_2(\Pr(y_1^4|u_1^4))$ when the frozen-location pattern is `$\mathcal{FDDD}$'.}
\label{fig:divide_conquer_SC}
\end{figure*}


Under Assumption 1, considering Eq.~\eqref{eq:MLD_SC_nothing}, if $\Omega_{01}^{(j)}=\varnothing$,
the maximal value of $T$ is just a product of the maximal value of $T_1$ and the maximal value of
$T_2$. For example, Fig.~\ref{fig:divide_conquer_SC}(b) shows an example of frozen-location pattern
`$\mathcal{DDDD}$' which has $\Omega_{01}^{(0)}=\varnothing$, when $M=4$ and
$N=4$. If $\Omega_{01}^{(j)}$ is not empty, for any $i \in \Omega_{01}^{(j)}$,
$v_{\frac{jM}{2}+i}=u_{jM+2i}$.

%
%
%

\subsection{Divide-and-conquer AML decoding unit for SCL-based algorithms}
Extending the idea in Eqs.~\eqref{eq:MLD_SC_nothing} and \eqref{eq:MLD_SC}, we propose a divide-and-conquer AML decoding
method for SCL-based algorithms under Assumption 1. For SC-based algorithms,
only the maximal value of ${\Pr}(\mathbf{y},\hat{u}_1^{jM}|u_{jM+1}^{jM+M})$ is
needed. In contrast, for SCL-based algorithms with list size $L$, the $L$
maximal values of ${\Pr}(\mathbf{y},\hat{u}_1^{jM}|u_{jM+1}^{jM+M})$ are
needed. A simple understanding for our method is that the
$\max(\Pr(\rho))$ function is replaced by a function finding
the $L$ maximal values of $\Pr(\rho)$, denoted as
$[\Pr(\rho_1),\cdots,\Pr(\rho_L)]={\rm{max}}_L(\Pr(\rho))$. ${\rm{max}}_L(\Pr(\rho))\boxtimes{\rm{max}}_L(\Pr(\psi))$
generates $L^2$ values of $\Pr(\rho_i)\Pr(\psi_j)$ for $1 \leq i,j \leq
L$. 

The path expansion-and-pruning procedure of SCL-based algorithms is divided into
two stages. In the first stage, the $q$ most reliable paths are selected for each list by
calculating and comparing path metrics. In the second stage, the $L$ most reliable paths among the
$qL$ survival paths generated in the first stage. This two stage approach was
proposed in our prior work \cite{SBSCL2014SiPS}, and the novelty herein is that we use the divide-and-conquer method
to reduce the complexity of the first stage. The second stage has been
described in \cite{SBSCL2014SiPS}, and we omit its discussions.

%
%
%
%
%
%
%
%
%
%
%

Assuming $\lvert \Omega_{01}^{(j)} \rvert = \beta_j$,
$\lvert\Omega_{11}^{(j)}\rvert = \gamma_j$. When $\beta_j \geq 1$, let
$\Omega_{01}^{(j)}=\{i_1^{(j)},i_2^{(j)},\cdots,i_{\beta_j}^{(j)}\}$ $(i_1^{(j)} < i_2^{(j)} <\cdots <i_{\beta_j}^{(j)})$. The first stage includes:

%
%
%

Step 0: the RCC method \cite{SBSCL2014SiPS} is applied to calculate both $T_1$
  and $T_2$.

Step 1: Given any $\beta_j$-bit
  binary vector $\mathcal{B}^{(j)}=(u_{jM+2i_1^{(j)}},u_{jM+2i_2^{(j)}},\dots,u_{jM+2i_{\beta_j}^{(j)}})$, there are $2^{\gamma_j}$
  possible values for both $v_{\frac{jM}{2}+1}^{\frac{jM+M}{2}}$ and $u_{jM+1,e}^{jM+M}$.
  We find the $\min(q,2^{\gamma_j})$ maximal values of $2^{\gamma_j}$ values of
  $T_1$,
  and the $\min(q,2^{\gamma_j})$ maximal values of $2^{\gamma_j}$ values of
  $T_2$.

Step 2: For $\mathcal{B}^{(j)}$, there are
  $\bigl(\min(q,2^{\gamma_j})\bigr)^2$ values of $T$, which
  is a product of values of $T_1$
  and $T_2$ generated in
  Step 1. 

Step 3: The $q$ maximal values are selected from $\bigl(\min(q,2^{\gamma_j})\bigr)^22^{\beta_j}$ values of
  $T$ generated by Step 2 because
  there are $2^{\beta_j}$ possible values for $\mathcal{B}^{(j)}$.

If $\Omega_{01}^{(j)}= \varnothing$ and $\beta_j =0$, we still use the
aforementioned four steps to find the $q$ most reliable paths for each list except that $\mathcal{B}^{(j)}$ is
considered as a void binary vector which is the only value for $\mathcal{B}^{(j)}$
when $\beta_j =0$.

Fig.~\ref{fig:divide_conquer_SC}(d) and \ref{fig:divide_conquer_SC}(e) show two
examples for frozen-location patterns `$\mathcal{DDDD}$' and `$\mathcal{FDDD}$',
respectively when $M=4$, $N=4$, and $q=2$.
After these four steps are carried out for each list, there are $qL$ values of
$T$ left, which are sorted to
choose the $L$ maximal values in the second stage.

The proposed divide-and-conquer AML decoding unit has a lower computational
complexity. It reduces the number of symbol-wise channel transition probabilities dealt by the
list pruning function by sorting the intermediate calculation results generated
by the RCC method \cite{SBSCL2014SiPS}, whereas the DMM, RCC, and DRH methods perform list pruning function after all the symbol-wise
channel transition probabilities are calculated. For example, in
Fig.~\ref{fig:divide_conquer_SC}(d), the DMM, RCC, and DRH methods perform
${\rm{max}}_2(\Pr(y_1^4|u_1^4))$ after all 16 values of $\Pr(y_1^4|u_1^4)$ are
calculated. The proposed divide-and-conquer method performs ${\rm{max}}_2(\Pr(y_1^2|v_1^2))$ and ${\rm{max}}_2(\Pr(y_3^4|u_2,u_4))$ first. Then 
it finds the two maximal values out of four elements generated by
${\rm{max}}_2(\Pr(y_1^2|v_1^2))\boxtimes {\rm{max}}_2(\Pr(y_3^4|u_2,u_4))$.
The output of the proposed AML decoding unit is the same as those of other ML
decoding units if they have the same input.

Given an $M$-bit symbol $u_{jM+1}^{jM+M}$, $\Omega_{01}^{(j)}$,  $\Omega_{11}^{(j)}$,
${\Pr}(y_1^{\frac{N}{2}},\hat{v}_{1}^{\frac{jM}{2}}|v_{\frac{jM}{2}+1}^{\frac{jM+M}{2}})$,
and ${\Pr}(y_{\frac{N}{2}+1}^{N},\hat{u}_{1,e}^{jM}|u_{jM+1,e}^{jM+M})$, the first
stage using the divide-and-conquer decoding unit needs two $2^{\gamma_j}$-to-$\bigl(\min(q,2^{\gamma_j})\bigr)$
sorts, one $\bigl(\min(q,2^{\gamma_j})\bigr)^22^{\beta_j}$-to-$q$ sort, and
$\bigl(\min(q,2^{\gamma_j})\bigr)^22^{\beta_j}$ multiplications per list,
whereas the ML decoding unit in \cite{SBSCL2014SiPS} needs
$2^{\beta_j+2\gamma_j}$ multiplications and a $2^{\beta_j+2\gamma_j}$-to-$q$
sort per list. By examining all possible values of $\beta_j$ and $\gamma_j$, we
can find the worst-case computational complexity.

We demonstrate the advantage of the proposed divide-and-conquer AML unit in
computational complexity as opposed
to other ML decoding units with an example of $M=8$ and
$q=4$. Henceforth, we only discuss the computational complexity per list to accomplish the job of
the first stage of the proposed method. 
Table~\ref{tab:Comp_complexity} lists worse-case computational complexities of
different methods and shows that the proposed method has the smallest
computational complexity when 81 eight-bit frozen-location patterns under
Assumption 1 are needed to be dealt with.

\begin{table}[hbtp]
\begin{center}
\begin{threeparttable}[b]
\caption{Worst-case computational complexity of different methods when $M=8$ and $q=4$.}
\label{tab:Comp_complexity}
\begin{tabular}{|c|l|}
\hline
 Method & Computational Complexity\\ \hline
RCC \cite{SDSCL} \tnote{$\ddagger$} & 304 multiplications, a 256-to-4 sort \\ \hline
DMM \cite{6464502} \tnote{$\ddagger$} & 1792 multiplications, a 256-to-4 sort \\ \hline
DRH \cite{reduced_latency_VLSI} \tnote{$\ddagger$} & 784 multiplications, a 256-to-4 sort \\ \hline
Divide-and-Conquer & 112 multiplications, a 64-to-4 sort\\ 
AML \tnote{$\ddagger$} & and two 16-to-4 sorts \\ \hline
Divide-and-Conquer & 80 multiplications, a 32-to-4 sort \\ 
AML \tnote{$\dagger$} &  and two 16-to-4 sorts\\ \hline\hline
LC-AML \tnote{$\star$} & 80 multiplications, a 32-to-4 sort \\ 
& and four 8-to-4 sorts \\ \hline
\end{tabular}
\begin{tablenotes}
\footnotesize
\item [$\ddagger$] All 81 eight-bit patterns under
  Assumption 1 are dealt with.
\item [$\dagger$] Only nine eight-bit patterns in
  Sec.~\ref{sec:flp_bec} are dealt with. 
\item [$\star$] Only six eight-bit patterns in
  Sec.~\ref{sec:ML_design} are dealt with. 
\end{tablenotes}
\end{threeparttable}
\end{center}
\end{table}


\begin{figure}[htbp]
\centering
\includegraphics[width=7.5cm]{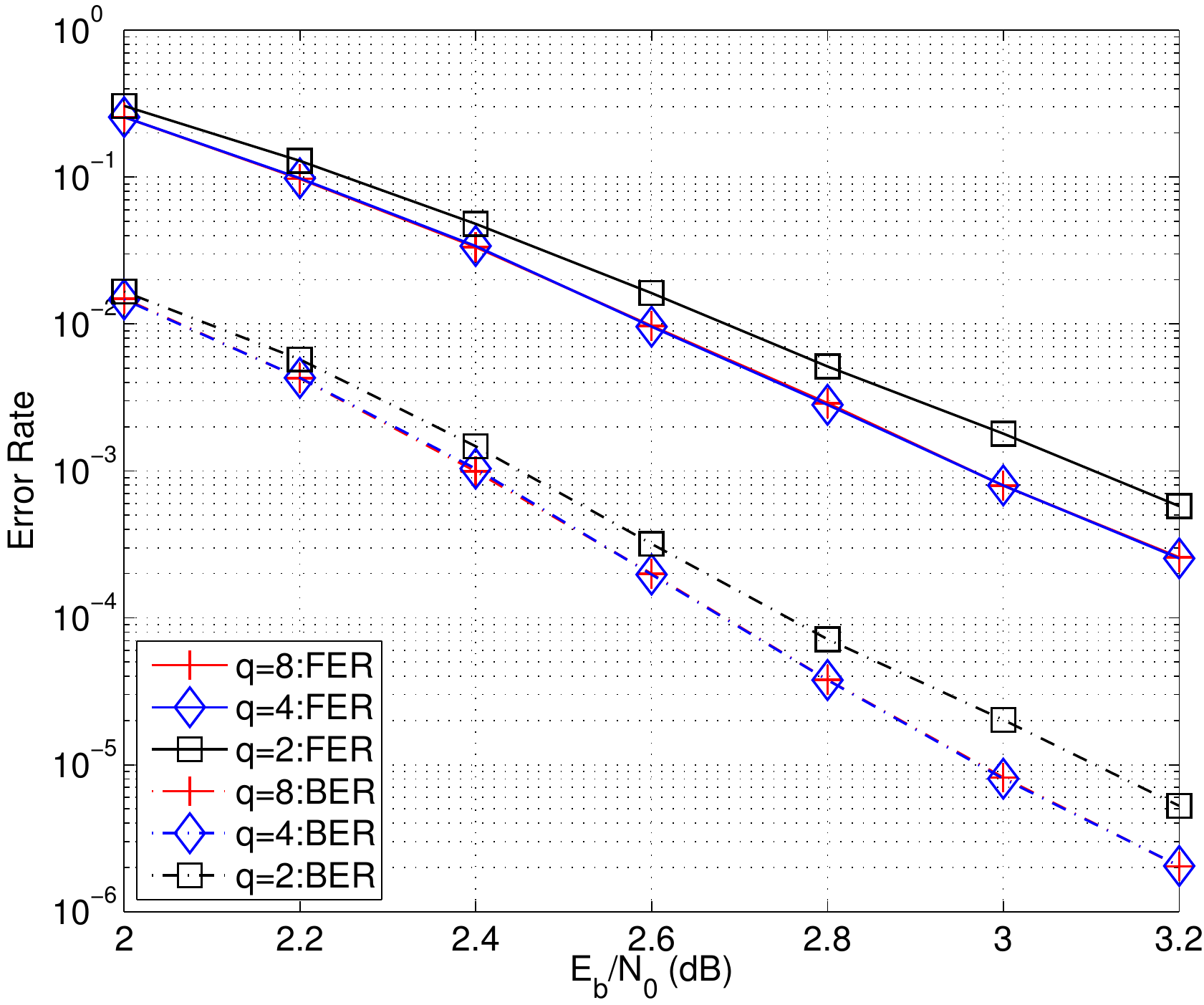}
\caption{Frame and bit error rates of CA-SCL decoders with different $q$s for a (2048,1433) polar
  codes with a 32-bit CRC over the AWGN channel when adapting the LC-AML decoding unit.}
\label{fig:AML_per}
\end{figure}

Regarding the impact on the error performance, our proposed method has the same
performance degradation as in \cite{SDSCL}. If $q\geq L$, our method does not introduce any
performance degradation for SCL-based algorithms. If $q < L$,  the performance degradation depends on
values of $q$ and $L$, and the performance degradation is usually negligible
when $q$ and $L$ are small. Fig.~\ref{fig:AML_per} shows the frame and bit error rates of a
(2048, 1433) polar code with a 32-bit CRC of CA-SCL decoders with the LC-AML decoding unit when $M=8$ and
$L=8$. When $q=4$, the performance loss is negligible. However, $q=2$ leads to a
performance loss of about 0.1 dB at an FER level of $10^{-3}$.

\section{Frozen-Location Patterns for Polar codes}
\label{sec:data_pattern}

Considering the hardware implementation for the divide-and-conquer AML unit, a uniform hardware design for all frozen-location patterns is preferred rather than different dedicated designs for various frozen-location
patterns. For $M=8$ and $M=16$, there are $81$ and $6561$ possible frozen-location patterns satisfying Assumption
1, respectively. Actually, some of them may never exist in a polar
code. Therefore, we want to know the exact number of frozen-location patterns in
a polar code, since the
number of frozen-location patterns impacts the complexity of
the divide-and-conquer AML decoding unit for SCL-based algorithms: the more frozen-location patterns,
the higher complexity the divide-and-conquer AML decoding unit has.

\subsection{Polar Codes for the BEC}
\label{sec:flp_bec}
\begin{figure}[htbp]
\centering
\includegraphics[width=8cm]{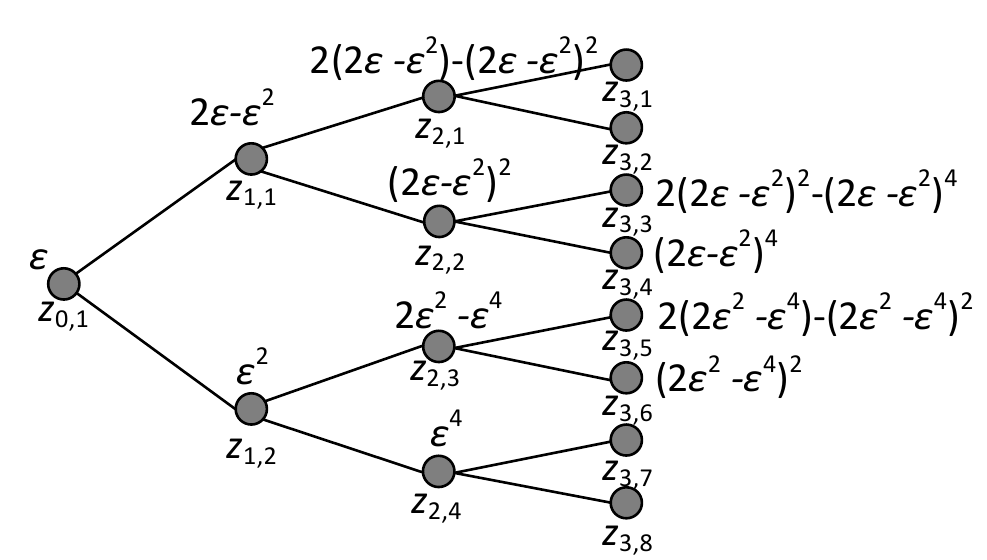}
\caption{Recursive calculation of erasure probabilities for polar codes over the
  BEC when the code length is no more than eight.}
\label{fig:Erasure_prob_sym8}
\end{figure}

For polar codes constructed for the BEC with an erasure probability
$\epsilon\text{ }(0 < \epsilon < 1)$, Eqs.~\eqref{eq:BEC_cons} and
\eqref{eq:BEC_cons2} are used in
\cite{4542778}. Fig.~\ref{fig:Erasure_prob_sym8} illustrates the transition
graph of the erasure 
probability for constructing a polar code with the code length no more than eight, and also can be
viewed as a sub-graph of the erasure probability transition for any eight-bit
symbol of a polar code because of the recursive calculation. In order to examine
frozen-location patterns in a polar code, we have following results
regarding the ordering of $z_{i,j}$ for $i\geq 1$ and $ 1 \leq j \leq
2^i$. This ordering determines possible frozen-location patterns in a polar code.

\begin{PROP}
\label{prop:1}
Assuming $z_{0,1}=\epsilon \in (0,1)$, given any $i \geq 1$ and $1\leq j \leq
2^i$, $z_{i,j}$ is calculated by Eqs.~\eqref{eq:BEC_cons} or
\eqref{eq:BEC_cons2}. We have
\begin{enumerate}[label=(\alph*)]
\item $0<z_{i,j}<1$ for $i \geq 1$ and $1\leq j \leq 2^i$,
\item $z_{i,2j-1} > z_{i,2j}$ for $i \geq 1$ and $1\leq j \leq 2^{i-1}$,
\item $z_{i,4j-3} > z_{i,4j-2} > z_{i,4j-1} > z_{i,4j}$ for $i \geq 2$ and
  $1\leq j \leq 2^{i-2}$,
\item $z_{i,8j-7} > z_{i,8j-6} > z_{i,8j-5} > z_{i,8j-3} > z_{i,8j-4} >
z_{i,8j-2} > z_{i,8j-1} > z_{i,8j}$  for $i \geq 3$ and $1\leq j \leq
2^{i-3}$.
\end{enumerate}
\end{PROP}

Proof of Prop.~\ref{prop:1} is provided in the Appendix.

Now, let us explain how the ordering of $z_{n,j}$ determines $2^m$-bit ($1\leq m
\leq 3$) frozen-location patterns in an ($N, K$) polar code over the
BEC. First, to choose elements of $\mathcal{A}^c$ for an ($N, K$) 
polar code over the BEC, $\mathcal{A}^c$ is chosen such that $\sum_{j\in\mathcal{A}^c}z_{n,j}$
is maximal and $\lvert \mathcal{A}^c \rvert = N-K$, where $N=2^n$. Then, if
there are $k_j$ frozen bits in a symbol $u_{2^mj+1}^{2^m(j+1)}$ $(0\leq j<
\frac{N}{2^m})$, a set $\mathcal{A}^c_j$ consisting of indexes of these $k_j$
frozen bits must be chosen such that $\sum_{t\in\mathcal{A}^c_j}z_{n,t}$ is
maximal while $\lvert{A}^c_j\rvert=k_j$. For example, assuming there are four
frozen bits in $u_1^8$ in a (16, 12) polar code, by Proposition \ref{prop:1}(d), $z_{4,1} > z_{4,2} > z_{4,3} > z_{4,5} > z_{4,4} >
z_{4,6} > z_{4,7} > z_{4,8}$. Hence, $u_1, u_2, u_3$, and $u_5$ will be frozen
bits and the frozen-location pattern for $u_1^8$ will
be `$\mathcal{FFFDFDDD}$'.

Therefore, for polar codes constructed by the method in \cite{4542778}, by
Proposition \ref{prop:1}(b), there are three two-bit frozen-location patterns:
`$\mathcal{DD}$', `$\mathcal{FD}$', and `$\mathcal{FF}$'. We note that the
implication of Proposition \ref{prop:1}(b) is the counterpart over the BEC
of Assumption 1 in \cite{ParSC3}. By Proposition
\ref{prop:1}(c), there are five four-bit frozen-location patterns:
`$\mathcal{DDDD}$', `$\mathcal{FDDD}$', `$\mathcal{FFDD}$', `$\mathcal{FFFD}$',
and `$\mathcal{FFFF}$'. By Proposition \ref{prop:1}(d), there are nine eight-bit
frozen-location patterns: `$\mathcal{DDDDDDDD}$', `$\mathcal{FDDDDDDD}$',
`$\mathcal{FFDDDDDD}$', `$\mathcal{FFFDDDDD}$', `$\mathcal{FFFDFDDD}$',
`$\mathcal{FFFFFDDD}$', `$\mathcal{FFFFFFDD}$', `$\mathcal{FFFFFFFD}$',
and `$\mathcal{FFFFFFFF}$'.

For a larger symbol size, it is hard to get the ordering of $z_{i,j}$
by an analytical method. A numerical method can be used. For example, the symbol size
is 16. 
By Proposition \ref{prop:1}(d), we have $z_{4,1} > z_{4,2} > z_{4,3} > z_{4,5} > z_{4,4} > z_{4,6} >
z_{4,7} > z_{4,8}$ and $z_{4,9} > z_{4,10} > z_{4,11} > z_{4,13} > z_{4,12} > z_{4,14} >
z_{4,15} > z_{4,16}$. We also have $z_{4,5} > z_{4,9} > z_{4,7} >
z_{4,11}$ and $z_{4,4} > z_{4,6} > z_{4,10} > z_{4,8} >
z_{4,12}$. For $0 < z_{0,1}=\epsilon < 1$, 
\[z_{4,10}-z_{4,7} =
2\epsilon^4(\epsilon-1)^4\left[\epsilon^3(\epsilon^2-\epsilon+24)(\epsilon-1)^3-8)\right]
< 0.\]

Moreover, for $0 < z_{0,1} = \epsilon < 1$, $z_{4,4}-z_{4,9} =
2\epsilon^2(\epsilon-1)^4(\epsilon^{10}-4\epsilon^9+34\epsilon^8-116\epsilon^7+237\epsilon^6-375\epsilon^5+420\epsilon^4-280\epsilon^3+102\epsilon^2-16\epsilon-4)
<
0$ and
$z_{4,8}-z_{4,13}=2\epsilon^4(\epsilon-1)^2(\epsilon^{10}-6\epsilon^9+43\epsilon^8-132\epsilon^7+251\epsilon^6-262\epsilon^5+121\epsilon^4-8\epsilon^3-6\epsilon^2-4\epsilon-2)
< 0$. These two inequalities are verified numerically.

Because of the recursive calculation of $z_{i,j}$, for $i\geq 4$ and $1 \leq j \leq
2^{i-4}$, we have
\begin{equation*}
\begin{split}
z_{i,16j-15} &> z_{i,16j-14} > z_{i,16j-13} > z_{i,16j-11} \\
&> z_{i,16j-7} > z_{i,16j-12} > z_{i,16j-10} > z_{i,16j-9} \\
&> z_{i,16j-6} > z_{i,16j-5} > z_{i,16j-3} > z_{i,16j-8} \\
&> z_{i,16j-4} > z_{i,16j-2} > z_{i,16j-1} > z_{i,16j}.
\end{split}
\end{equation*}
Thus, there are only 17 frozen-location patterns for 16-bit symbols.

It is not meaningful to consider the symbol size greater than 16, because this will
incur very high complexity for hardware implementations. 


\subsection{Polar Codes for the AWGN Channel}
\label{sec:flp_awgn}
For the construction method introduced in \cite{6823688} for the AWGN channel,
it is difficult to analyze the relationship between $z_{3,i}$'s
for $1 \leq i \leq 8$ based on Eqs. \eqref{eq:AWGN_construction} and
\eqref{eq:phi_func}. Instead, we examine eight polar codes constructed with the method in \cite{6823688}, which have code
lengths from $2^{10}$ to $2^{13}$ and code rates of 0.5 and 0.8 to identify eight-bit
frozen-location patterns. By examining all
eight-bit symbols of these polar codes, we found that in these
codes there are only nine eight-bit frozen-location patterns, which are the same as those for
polar codes constructed for the BEC, listed in Sec.~\ref{sec:flp_bec}. Our
observation is consistent with Assumption 1 in \cite{ParSC3}. 




\subsection{Computational Complexity of the Divide-and-Conquer AML Decoding Unit}

When it needs to deal with only the frozen-location patterns mentioned in Sections \ref{sec:flp_bec} and
\ref{sec:flp_awgn}, the divide-and-conquer AML decoding
unit has a smaller complexity. If $M=8$ and $q=4$, it needs 80 multiplications, two 16-to-4
sorts, and a 32-to-4 sort, as listed in Table~\ref{tab:Comp_complexity}. It saves 32 multiplications, a 32-to-4
sort and a 8-to-4 sort compared with the divide-and-conquer AML decoding
unit which deals with all 81 frozen-location patterns following Assumption 1,
since a 64-to-4 sort consists of two 32-to-4 sorts and a 8-to-4 sort. 

If $M=16$ and $q=4$, to deal with all $3^8=6561$ 16-bit frozen-location patterns
satisfying Assumption 1, the first stage of the proposed ML decoding unit needs 1632
multiplications, two 256-to-4 sorts, and a 1024-to-4 sort. However, to deal with
17 16-bit frozen-location patterns discussed in Section \ref{sec:flp_bec}, the simplified
divide-and-conquer AML decoding unit needs 736 multiplications, two 256-to-4
sorts, and a 128-to-4 sort.

\section{Low-Complexity AML Decoding Unit}
\label{sec:ML_design}
For convenience, we implement the proposed divide-and-conquer AML decoding unit, assuming
$M=8$ henceforth. Our implementation can be readily extended to other values of
$M$. To further reduce complexity and latency, we do not use the
divide-and-conquer method to deal with patterns `$\mathcal{DDDDDDDD}$',
`$\mathcal{FFFFFFFD}$', and `$\mathcal{FFFFFFFF}$', which will be described in
Sec.~\ref{subsec:ae_scl}. Then the divide-and-conquer AML decoding unit can be simplified further by
dealing with only the remaining six eight-bit frozen-location patterns. This
simplified divide-and-conquer AML decoding unit, referred to as the LC-AML
decoding unit, needs 80 multiplications, four 8-to-4 sorts, and a 32-to-4
sort. It saves two 8-to-4 sorts compared with the divide-and-conquer AML
decoding unit dealing with nine patterns, since a 16-to-4 sort consists of three
8-to-4 sorts. This also leads to a shorter critical path in our design than the
divide-and-conquer AML decoding unit.



%

\subsection{Hardware Design for the LC-AML Decoding Unit}
SCL-based polar decoders in the literature can be divided into two categories:
the log-likelihood (LL) based decoders \cite{JunPolarList, 6865312, SDSCL} and the log-likelihood-ratio (LLR) based
decoders \cite{7114328, reduced_latency_VLSI}. Although our proposed algorithm in Sec.~\ref{sec:LC_MLD_SCL} is described in the probability
domain, it can be easily adapted for both the LL-based decoder and the LLR-based
decoder. We focus on the LLR-based polar decoder, because in general the
LLR-based decoder has a better area efficiency than the LL-based decoder.

First, we adapt the proposed LC-AML decoding unit to the LLR-based SCL
decoder. Given path metrics $\text{PM}_{k}^{(t)}$ of $L$ list survivors and
assuming $u_t$ 
is the last bit processed by the decoder, where $1 \leq k\leq L$, $1\leq t\leq
N$, and $t$ is a multiple of $M$. Suppose $\alpha_{j,l}$ $(0 \leq j < M)$
represents the LLR of
${\Pr}(y_{j\frac{N}{M}+1}^{(j+1)\frac{N}{M}},\hat{w}_{1+j\frac{N}{M}}^{\frac{t}{M}+j\frac{N}{M}}|w_{\frac{t}{M}+1+j\frac{N}{M}})$
corresponding to the list $l$. The path metric $\mathcal{PM}_{k,p}^{(t+M)}$
of the $p$-th expanded path from the $k$-th list survivor corresponding to
$u_{t+1}^{t+M}=p$ $(0 \leq p < 2^M)$ is $\mathcal{PM}_{k,p}^{(t+M)} = \text{PM}_{k}^{(t)} + \sum_{j=0}^{M-1}m_j\lvert\alpha_{j,l} \rvert,$
where $m_j=0$ if
$w_{\frac{t}{M}+1+j\frac{N}{M}}=\frac{1}{2}(1-\text{sign}(\alpha_{j,l}))$
\cite{7114328}. Otherwise, $m_j=1$. Then our goal is to calculate
$\mathcal{PM}_{k,p}^{(t+M)}$ and to select the $L$ minimum
values of $\mathcal{PM}_{k,p}^{(t+M)}$. 

\begin{figure}[htbp]
\centering
\includegraphics[width=8cm]{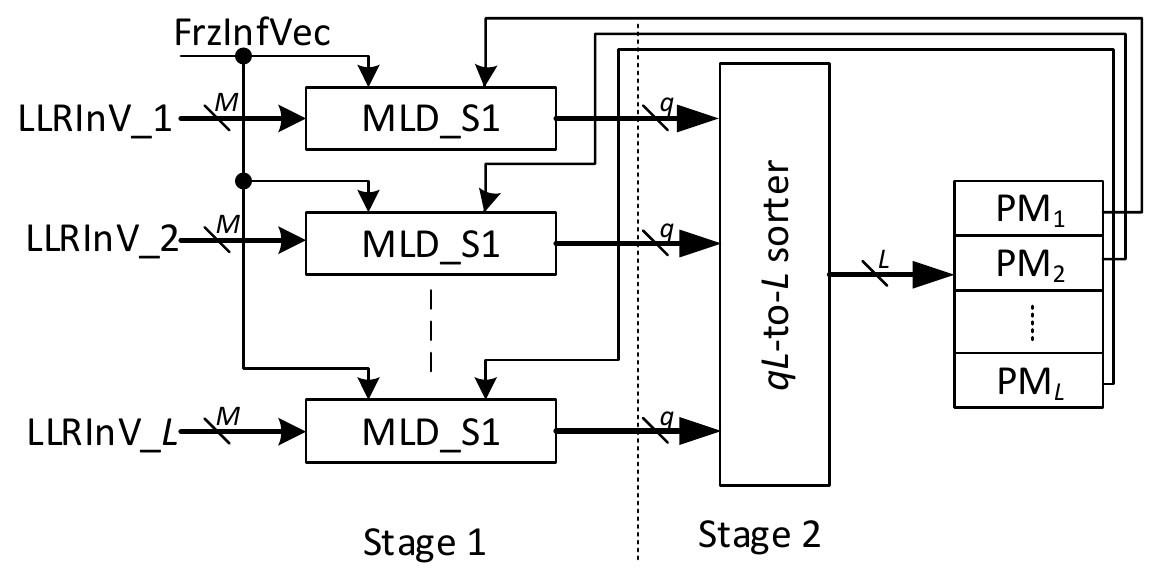}
\caption{Top architecture of the proposed LC-AML decoding unit.}
\label{fig:top_arch_MLD}
\end{figure}

Fig.~\ref{fig:top_arch_MLD} shows the top architecture of our low-complexity implementation
for the LC-AML decoding unit. MLD\_S1 calculates path metrics
and selects the $q$ minimum values for each list. FrzInfVec is an $M$-bit frozen
bit indication vector $(f_1,f_2,\cdots,f_M)$ for $u_{t+1}^{t+M}$. For $1\leq j
\leq M$, if $u_{t+j}$ is
a frozen bit, $f_j=1$; otherwise, $f_j=0$. LLRInV\_$l$ is the vector
$(\alpha_{0,l},\alpha_{1,l},\cdots,\alpha_{M-1,l})$ for $1 \leq l \leq L$. 

\begin{figure*}[htbp]
\centering
\includegraphics[width=18cm]{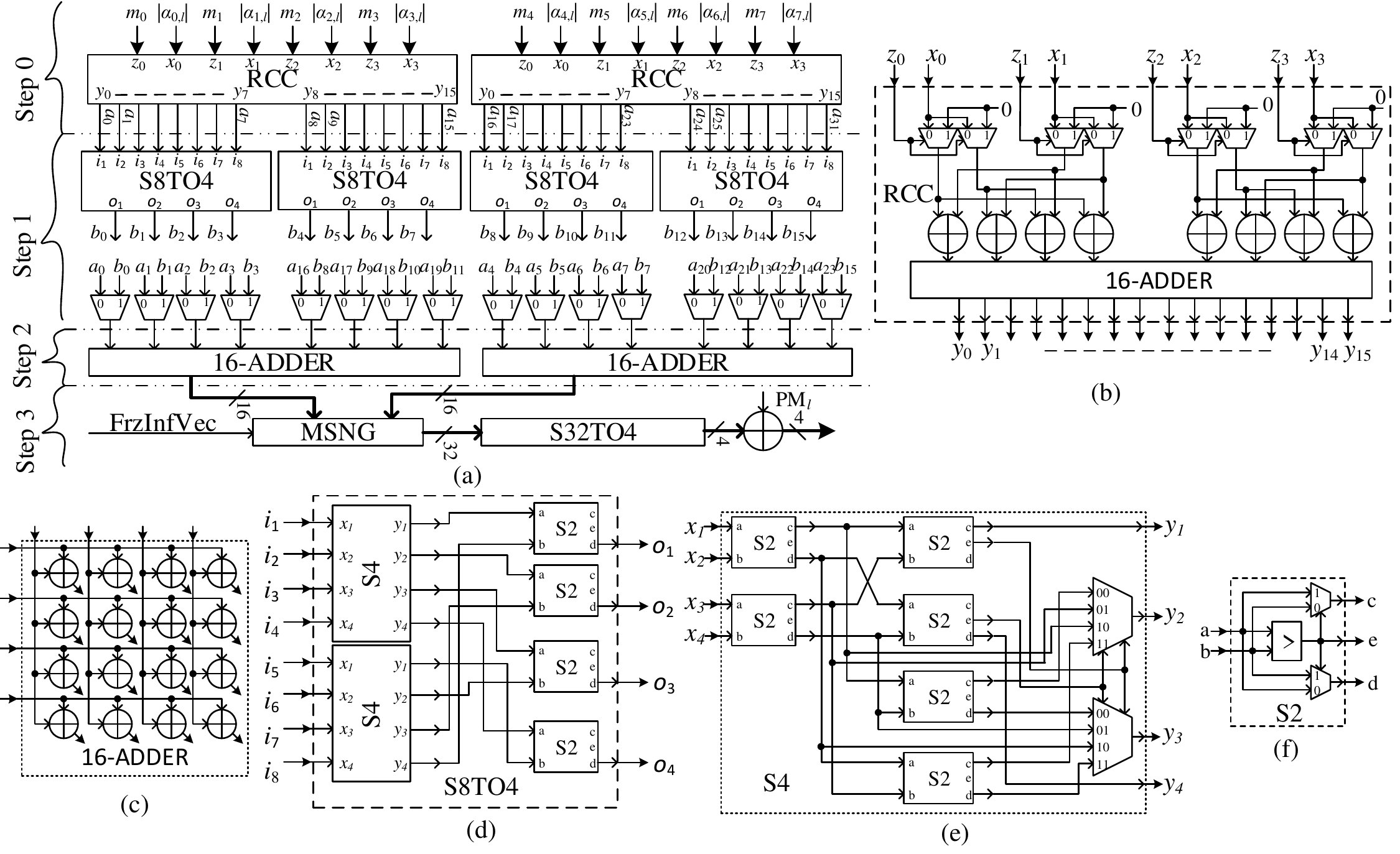}
\caption{Design of MLD\_S1\_$q4$ when $M=8$ and $q=4$. (a) top architecture of
  MLD\_S1\_$q4$, (b) diagram of RCC, (c) diagram of 16-ADDER, (d)
  diagram of S8TO4, (e) diagram of S4, (f) diagram of S2.}
\label{fig:MLD_S1}
\end{figure*}

Fig.~\ref{fig:MLD_S1}(a) shows the design for MLD\_S1\_$q4$ when $M=8$ and
$q=4$. Here, we focus on the data path for calculating path metrics. The circuitry to generate
symbol values associated with path metrics is simple and consists of
XORs, and therefore is omitted. 
The data paths corresponding to different
steps aforementioned in Section \ref{sec:LC_MLD_SCL} are labeled as well.  

In Step 0, two RCC blocks, shown in
Fig.~\ref{fig:MLD_S1}(b), are used. LLR $a_i$ ($16 \leq i\leq 31$) associated with
${\Pr}(y_{\frac{N}{2}+1}^{N},\hat{u}_{1,e}^{8j}|u_{8j+1,e}^{8j+8}=(i-16)_2)$ is
calculated by the right RCC block. LLR $a_i$ $(0 \leq i \leq 15)$ associated with ${\Pr}(y_1^{\frac{N}{2}},\hat{v}_{1}^{4j}|v_{4j+1}^{4j+4}=(i)_2)$ is
calculated by the left RCC block. Here, $(i)_2$ represents the binary string of
interger $i$. 16-ADDER contains 16 adders to calculate path
metrics, shown in Fig.~\ref{fig:MLD_S1}(c).

In Step 1, for different frozen-location patterns, path metrics go through different data
paths selected by 16 2-to-1 multiplexers. Their control words are 1 if
frozen-location patterns are $\mathcal{FDDDDDDD}$, $\mathcal{FFDDDDDD}$;
otherwise they are 0.


In Step 2, results from Step 1 are combined to calculate $\sum_{j=0}^{7}m_j\lvert
\alpha_{j,l} \rvert$. 

In Step 3, there are 32 path metrics going through a 32-to-4 sorter. However,
for some frozen-location patterns, the number of valid symbol values is less than 32 because
the number of frozen bits can be larger than three. Therefore, path
metrics associated with those invalid symbol values need to be set to the
maximal positive value as well so that the
$4$ minimum path metrics belong to valid symbol values. MSNG accomplishes
this job with FrzInfVec, which contains the frozen-location pattern information.

Different sorters used in our design are shown in Fig.~\ref{fig:MLD_S1}(d), 5(e)
and 5(f). S8TO4 finds the minimum four values of eight values. S4 sorts the four
inputs and outputs them in decreasing order and has a shorter critical path of
two comparators and one 4-to-1 multiplexer than a four-input bitonic sorter
\cite{BatcherSorter}, which has a critical path of three comparators. S32To4
consists of seven S8TO4 units in a binary tree structure.

Although MLD\_S1\_$q4$ is designed for six eight-bit frozen-location patterns,
other frozen-location patterns also can be dealt with by MLD\_S1\_$q4$, such 
as all frozen-location patterns satisfying the following two conditions. First, the frozen-location
pattern has at least three `$\mathcal{F}$'s. Second, two frozen bits are located at
the first two bits of the data symbol. 


\subsection{Area-Efficient SCL Decoder}
\label{subsec:ae_scl}
To examine the advantage of our proposed design, we incorporate MLD\_S1\_$q4$
into CA-SCL polar decoders with list size $L=4$. Architecture-wise, our decoder, referred to as the
AE-SCL decoder, is almost the same as the architecture of the tree based
reduced latency SCL polar decoder in \cite{reduced_latency_VLSI}, which performs
the CA-SCL decoding algorithm on a binary tree representation of a polar
code, except that our AE-SCL decoder uses the LC-AML decoding unit instead of
the DRH ML decoding unit used in \cite{reduced_latency_VLSI}.

Leaf nodes of the decoding tree for our decoder are divided
into four categories: 
\begin{enumerate}
\item Rate-0 node: its frozen-location pattern contains only `$\mathcal{F}$', i.e., the
  node contains only frozen bits.
\item Rate-1 node: its frozen-location pattern contains only `$\mathcal{D}$', i.e., the
  node contains only information bits.
\item Repetition node\cite{6804939}: its frozen-location pattern is either `$\mathcal{FFFFFFFF\_FFFFFFFD}$'
  or  `$\mathcal{FFFFFFFD}$'.
\item Rate-R-2 node: its frozen-location pattern is one of the six eight-bit frozen-location patterns.
\end{enumerate}

Rate-0 and rate-1 nodes are decoded with the same methods as in
\cite{reduced_latency_VLSI}. The main difference between our proposed decoder
here and the tree based reduced latency SCL polar decoder is how to deal with
repetition nodes and rate-R-2 nods. For repetition nodes, a binary tree of adders
is used to calculate LLRs in order to reduce the decoding latency
\cite{6804939}. Rate-R-2 nodes are dealt with by MLD\_S1\_$q4$, which reduces the area of AE-SCL decoders.

\subsection{Synthesis Results}

AE-SCL decoders with $L=4$ are implemented for three polar codes: a (1024, 512) code, an
(8192,4096) code, and a (32768, 29504) code. These three codes are with a 32-bit
cyclic redundancy check. The number of processing units of
decoders for $N=1024$ is 256. For the other two codes, the decoder has 512
processing units. Five-bit channel LLRs are used. The synthesis tool is Cadence RTL compiler. The process
technology is TSMC 90nm CMOS technology. Here, four stages of pipeline registers are
used in the LC-AML decoding unit. Areas of different ML
decoding units for the (1024, 512) polar codes are listed in
Table~\ref{tab:syn_AML}. The area of our proposed LC-AML decoding unit is only one
fourth of that of the ML decoding unit in \cite{reduced_latency_VLSI}. By taking
into account fewer patterns, the area of the LC-AML decoding unit is 67\% of
that of the Divide-and-Conquer AML design which deals with all 81 eight-bit
frozen-location patterns following Assumption~1.  

\begin{table}[hbtp]
\begin{center}
\caption{Areas of different ML decoding units for the (1024,512) polar code.}
\label{tab:syn_AML}
\begin{threeparttable}[b]
\begin{tabular}{|c|c|c|c|}
\hline
& LC-AML\tnote{$\dagger$} & Divide-and-Conquer AML\tnote{$\ddagger$} &\cite{reduced_latency_VLSI} \\ \hline
area ($\text{mm}^2$) & 0.456 &0.673 & 2.298 \\ \hline
\end{tabular}
\begin{tablenotes}
\footnotesize
\item [$\dagger$] The LC-AML design targets the six eight-bit frozen-location patterns.
\item [$\ddagger$] The Divide-and-Conquer AML design targets all 81 eight-bit
  frozen-location patterns
  following Assumption 1.
\end{tablenotes}
\end{threeparttable}
\end{center}
\end{table} 

The synthesis results of three entire decoders
(AE-SCLs) are also listed in Tables~\ref{tab:Syn_result_1K},
\ref{tab:Syn_result_8K}, and \ref{tab:Syn_result_32K}, respectively. Here, NIT means the net information throughput. Compared
with decoders in \cite{reduced_latency_VLSI}, the SCL decoder architecture with
the best area efficiency to our knowledge, the AE-SCL decoders have smaller areas
because the proposed LC-AML decoding unit is applied. The
LC-AML decoding unit has  a slightly larger decoding
latency than that in \cite{reduced_latency_VLSI}, because the proposed LC-AML decoding unit deals with only eight-bit
frozen-location patterns, whereas the ML decoding unit in
\cite{reduced_latency_VLSI} can deal with some 16-bit frozen-location
patterns. Since the extra decoding cycles needed by AE-SCL decoders are a
very small fraction of the entire decoding cycles, the proposed
AE-SCL decoders still achieve better area efficiency than decoders in
\cite{reduced_latency_VLSI}. For example, for the (1024, 512) polar code, the
area efficiency of the AE-SCL decoder is 1.93 times of that of the decoder in
\cite{reduced_latency_VLSI}. As the code length increases, the advantage of area
efficiency is less because the ML decoding unit occupies a smaller fraction of
the entire decoder if the code is longer. Compared with symbol-decision SCL
decoders in \cite{SDSCL, 7114328, ParSC2}, the advantage of our decoders on
the area efficiency is more significant. The area efficiency of the AE-SCL
decoder is 3.32, 8.25, and 3.17 times of that of decoders in \cite{SDSCL,
  7114328, ParSC2}, respectively, for the (1024, 512) polar code.


\section{Multi-mode SCL Decoder}
\label{sec:mm_SCL}

All existing SCL polar decoders in the literature provide fixed throughput and
decoding latency given a polar code. These SCL decoders cannot adapt to variable
communication channels and applications. In order to adapt to different
throughput and latency requirements, we propose a multi-mode SCL (MM-SCL)
decoder with $n_d$ decoding paths, which can decode $P$ received words with list
size $L$ in parallel, where $1 \leq P, L \leq n_d$ and $n_d\geq P\times L$. This
multi-mode feature requires the decoder to perform SCL decoding algorithms with
different list sizes (the SC decoding algorithm is a special case of the SCL
decoding algorithm with list size $L=1$).



\subsection{Architecture Description}

\begin{figure}[htbp]
\centering
\includegraphics[width=9cm]{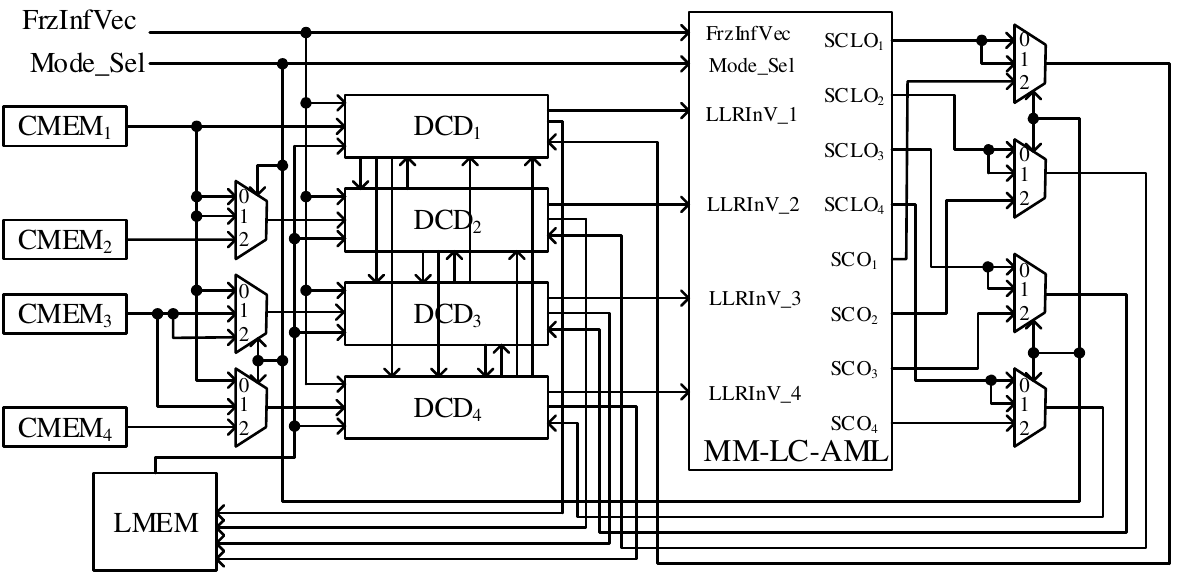}
\caption{Top architecture of the MM-SCL decoder when $n_d=4$.}
\label{fig:MM-SCL-top}
\end{figure}

 Assuming $n_d=4$, the top architecture of the MM-SCL decoder is shown in
Fig.~\ref{fig:MM-SCL-top}. It has four blocks of
channel memory, $\text{CMEM}_i$ $(1 \leq i \leq 4)$, to store four 
received codewords since the decoder of the MODE-1 mode can deal with four
received codewords simultaneously. Block $\text{DCD}_i$ $(1 \leq i \leq 4)$
contains processing units to calculate LLRs, and partial-sum units to 
update partial-sum for each list. The intermediate LLRs calculated by
$\text{DCD}_i$ are stored in block LMEM. Designs for processing units,
partial-sum units and the interface between processing units and LMEM adopt
blocks of the reduced-latency tree-based SCL decoder in
\cite{reduced_latency_VLSI}. We focus on the additional logic to support
multi-mode features. Mode\_Sel is a two-bit control word to
select the decoding mode of the MM-SCL decoder, shown in
Table~\ref{tab:mode_sel}.
\begin{table}[hbtp]
\begin{center}
\caption{Control word for the decoding mode of the MM-SCL decoder with $n_d=4$.}
\label{tab:mode_sel}
\begin{tabular}{|c|c|c|}
\hline
 Mode\_Sel & Decoding Mode & Notation\\ \hline
0 & SCL with $L=4$ & MODE-4\\ \hline
1 & SCL with $L=2$ & MODE-2\\ \hline
2 & SC & MODE-1\\ \hline
\end{tabular}
\end{center}
\end{table}

\begin{figure}[htbp]
\centering
\includegraphics[width=9cm]{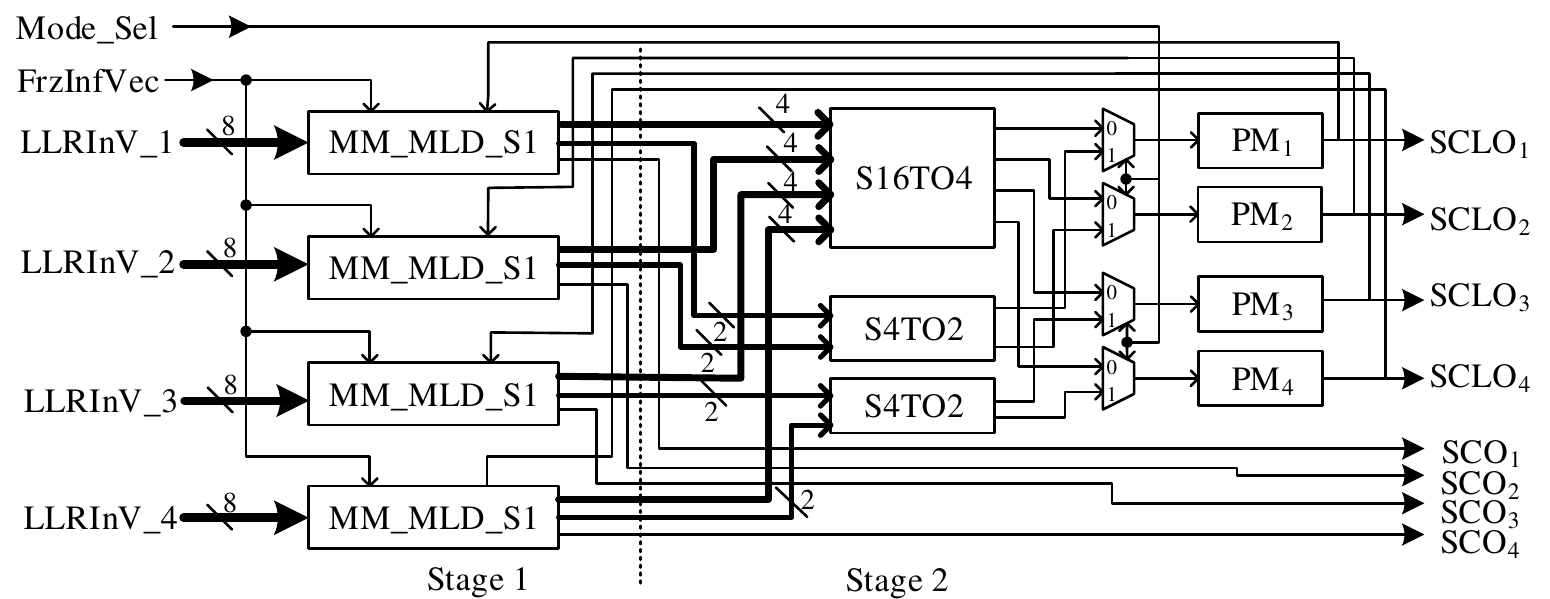}
\caption{Top architecture of MM-LC-AML for the MM-SCL decoder.}
\label{fig:top_mld_MM}
\end{figure}

Block MM-LC-AML performs the LC-AML decoding function for different types of leaf nodes
and is supposed to output the most reliable list candidate for the MODE-1 mode, the two
most reliable list candidates for the MODE-2 mode, and the four most reliable
list candidates for the MODE-4 mode.
The architecture in Fig.~\ref{fig:top_arch_MLD} is for a
fixed list size only. Here, we propose an MM-LC-AML unit (we take $n_d=4$ and $M=8$ as an
example) shown in Fig.~\ref{fig:top_mld_MM} to support the multi-mode
features. When the mode is MODE-4, all of $\text{SCLO}_1$, $\text{SCLO}_2$, 
$\text{SCLO}_3$, and $\text{SCLO}_4$ are used to decode a received codeword. When
the mode is MODE-2, $\text{SCLO}_1$ and $\text{SCLO}_2$ are used for one of two
received codewords; $\text{SCLO}_3$ and $\text{SCLO}_4$ are used for the other
received codeword. When the mode is MODE-1, each of $\text{SCO}_i (1\leq i \leq 4)$
is used by an individual received codeword.

Block MM\_MLD\_S1 performs the same function as block MLD\_S1\_$q4$ in
Fig.~\ref{fig:MLD_S1}, except that block MM\_MLD\_S1 supports multiple
modes. This can be accomplished by simply adding an S4 block between block S32TO4
and the adder at the bottom right of Fig.~\ref{fig:MLD_S1}(a). This
implementation has a disadvantage: the delay of the critical path is
increased due to the extra block in the data path and the decoding latency is
increased.

\begin{figure*}[htbp]
\centering
\includegraphics[width=12cm]{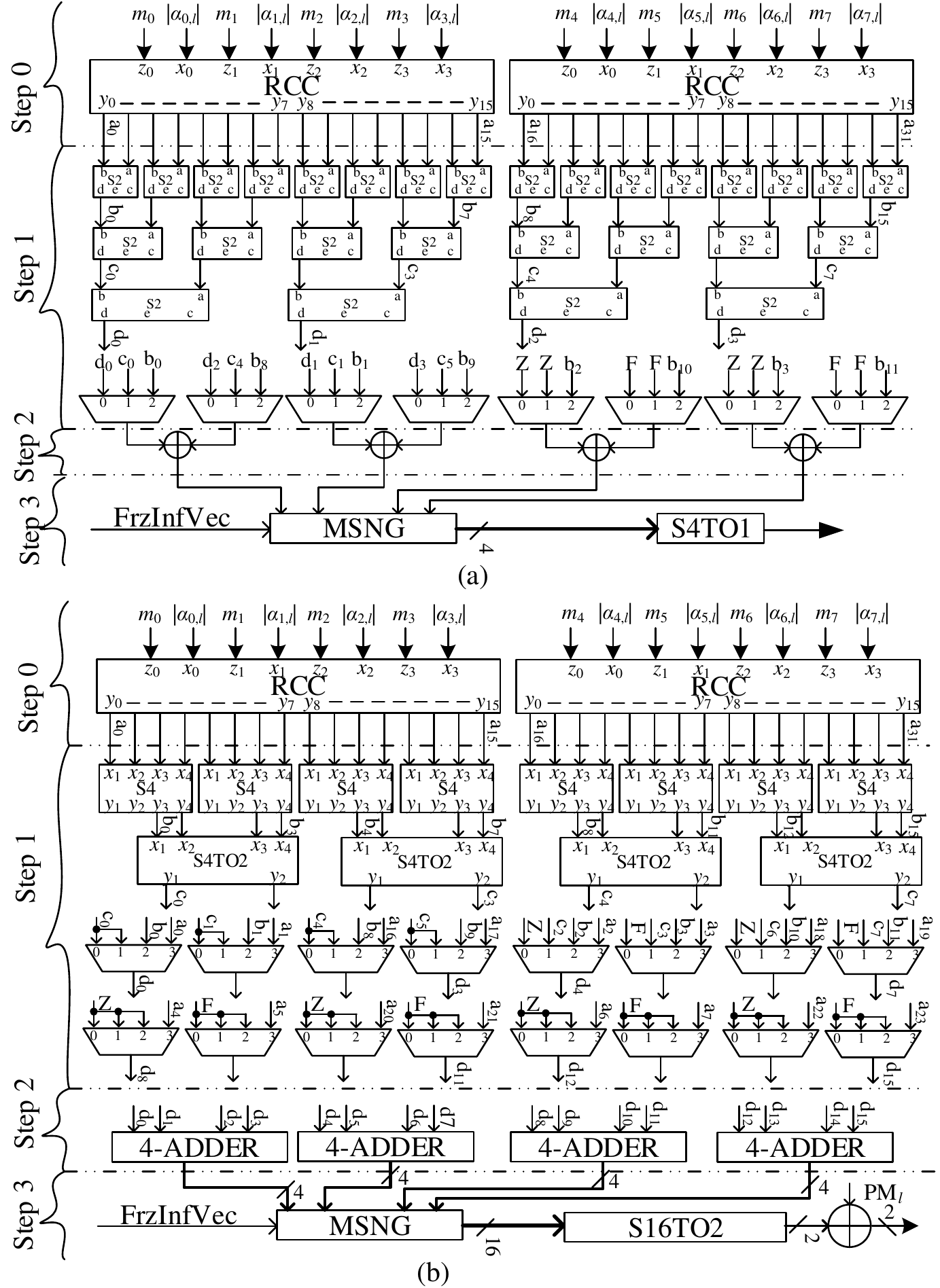}
\caption{(a) Design of MLD\_S1\_$q1$ for $q=1$, (b) design of MLD\_S1\_$q2$ for $q=2$.}
\label{fig:MLD_S1_L12}
\end{figure*}


To avoid the unnecessary increase of the decoding latency, we redesign the MLD\_S1
block for the MODE-1 and MODE-2 modes, respectively, called MLD\_S1\_$q1$ and
MLD\_S1\_$q2$, shown in Fig.~\ref{fig:MLD_S1_L12}(a) and \ref{fig:MLD_S1_L12}(b). Symbol values for `Z' and `F' are four-bit vectors
`0000' and `1111', respectively. Hence, the symbol value calculated from `Z' and
`F' is `11111111', which is guaranteed to be an invalid symbol value for our
designs. 

If MODE-2 is used, the control words for patterns `$\mathcal{FFDDDDDD}$',
`$\mathcal{FDDDDDDD}$', and `$\mathcal{FFFDDDDD}$' are 0, 1, and 2,
respectively; for the remaining patterns, the control words are 3. 
If MODE-1 is used, the control words for patterns `$\mathcal{FFDDDDDD}$',
`$\mathcal{FDDDDDDD}$', and `$\mathcal{FFFDDDDD}$' are 0, 0, and 1,
respectively; for the remaining patterns, the control words are 2.



Actually, MLD\_S1\_$q4$, MLD\_S1\_$q2$ and MLD\_S1\_$q1$ are integrated together
instead of three individual blocks in block MM\_MLD\_S1, since they have the same
circuitry for Step 0. Furthermore, sorting units of the top row of Step 1 in these three designs can
also be reused because block S8TO4 contains several S4 blocks and S2
blocks. The hardware sharing reduces the additional area for
supporting multiple modes and improves area efficiency without increasing the
critical path delay.

Compared with the AE-SCL decoder in Section~\ref{subsec:ae_scl}, the MM-SCL
decoder needs additional hardware for supporting multiple modes. The main
area increase is the additional three blocks of channel memories and the
hardware of MM\_MLD\_S1 to support the MODE-2 and MODE-1 modes.

\begin{figure}[htbp]
\centering
\includegraphics[width=9cm]{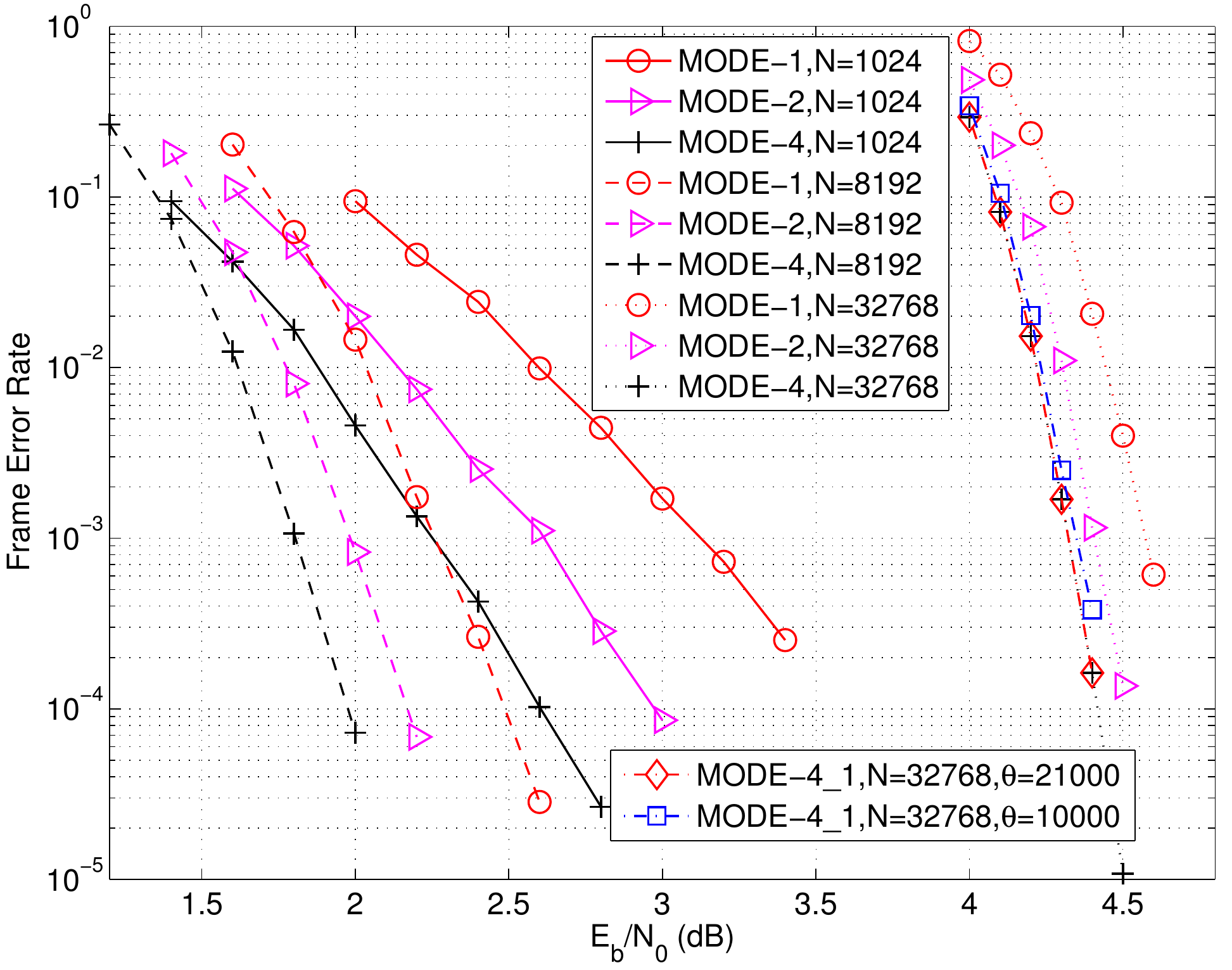}
\caption{Frame error rates of different modes for the MM-SCL decoder.}
\label{fig:MMSCL_all}
\end{figure}

Frame error rates of different modes for the MM-SCL decoder to decoder all the
three codes are shown in Fig.~\ref{fig:MMSCL_all}, which shows that, regarding
the frame error rates, MODE-4 $<$ MODE-2 $<$ MODE-1.

\subsection{Variation of modes}
An additional feature of the MM-SCL decoder is that Mode\_Sel can be changed during the decoding
procedure of a received word. This doe not need any additional hardware. It means that the
SCL algorithm with different list sizes can be used to deal with different
segments of a received word. For example, for a (32768, 29504) polar code,
$u_1^{21000}$ is decoded by the SCL algorithm 
of $L=4$ and $u_{21001}^{32768}$ is decoded by the SC algorithm. For the first
21000 bits, the decoder is in MODE-4 and four lists are maintained. The remaining bits are decoded by the SC
algorithm for each list. This mode is called MODE-4\_1. By choosing the
switching point $\theta$ (where the mode switches) of Mode\_Sel carefully, the decoding latency can be
improved slightly without any observed performance
loss as shown in Fig.~\ref{fig:MMSCL_all}. The decoding latency of the MODE-4 mode is
6718 cycles. MODE-4\_1 takes 6530 cycles to decode a received
codeword and improves the throughput slightly. To reduce the decoding latency
further, a smaller switching point can be used at the expense of small
performance loss. If $\theta=10000$, MODE-4\_1 has a decoding latency of  
6206 cycles and has a performance loss of about 0.03 dB compared with MODE-4 as shown in
Fig.~\ref{fig:MMSCL_all}, but still has a better performance than MODE-2. Hence,
MODE-4\_1 provides a more flexible way to achieve a decoding latency between
those of the MODE-4 mode and the MODE-1 mode. 

Therefore, the variation of modes provides a way for the MM-SCL decoder to
reduce decoding latency further somewhat without noticeable performance loss and
improves area efficiency further. It can also be used when decoding needs to be
finished as soon as possible due to external reasons, such as buffer overflow.


\subsection{Synthesis Results}

The MM-SCL decoder are implemented for the aforementioned three codes. For the
(1024, 512), the areas of the channel memory and the ML decoding unit are listed 
in Table~\ref{tab:overhead}. It shows that the
increased area of the MM-SCL decoder over the AE-SCL decoder is dominated by the
area of additional three blocks of channel memory. Due to the hardware sharing, the increased area of the
ML decoding unit is small.

\begin{table}[hbtp]
\begin{center}
\caption{Areas (in mm$^2$) of the channel memory and the ML decoding unit for MM-SCL and
  AE-SCL decoders when $N=1024$ and r=0.5.}
\label{tab:overhead}
\begin{tabular}{|c|c|c|c|}
\hline
  & MM-SCL & AE-SCL & Difference \\ \hline
Area of Channel Memory & 0.484 & 0.121 & 0.363 \\ \hline
Area of ML Decoding Unit  & 0.513 & 0.456 & 0.057 \\ \hline
\end{tabular}
\end{center}
\end{table} 

\begin{table*}[htb]
\begin{center}
\begin{threeparttable}[b]
\caption{Synthesizing results for different decoders when $N=1024$ and r=0.5.}
\label{tab:Syn_result_1K}
\begin{tabular}{|c||c||c|c|c||c|c|c|c|c|}
\hline
Decoder & AE-SCL & MODE-4 & MODE-2 & MODE-1 & \cite{reduced_latency_VLSI}
& \cite{SDSCL} & \cite{7114328} & \multicolumn{2}{c|}{\cite{ParSC2}} \\ \hline
List Size & 4 & 4 & 2 & 1 & \multicolumn{5}{c|}{4} \\ \hline
Area ($\text{mm}^2$) & 1.89 & \multicolumn{3}{c||}{2.31} & 3.83 & 1.70 & 1.78 & 2.14 & 4.10\tnote{*} \\ \hline
Clock Rate (MHz) & 409 & \multicolumn{3}{c||}{409} & 403 & 500 & 794 & 400 & 289\tnote{*} \\ \hline
\# of Decoding Cycles & 391 & 391 & 357 & 304 & 371 & 1540 & 2649 & \multicolumn{2}{c|}{1022} \\ \hline
Latency (us) & \LTP{391/409} & \LTP{391/409} & \LTP{357/409} & \LTP{304/409} &
\LTP{371/403} & \LTP{1540/500} & \LTP{2649/794} & \LTP{1022/400} & \LTP{1022/289}\tnote{*} \\ \hline
NIT (Mbps) & 547 & 547 & 1208 & 2887 & 570 & 155 & 154 & 200 & 144\tnote{*} \\ \hline
Area Eff. (Mbps/$\text{mm}^2$) & \AEP{547/1.89} & \AEP{547/2.31} & \AEP{1208/2.31} & \AEP{2887/2.31} & \AEP{570/3.83} & \AEP{155/1.70} & \AEP{154/1.78} & \AEP{200/2.14} &
\AEP{144/4.10}\tnote{*} \\ \hline
\end{tabular}
\begin{tablenotes}
\footnotesize
\item [*] Original synthesis results in \cite{ParSC2} are based on an ST 65nm CMOS
  technology. For a fair comparison, synthesis results scaled to a 90nm technology
  are used in the comparison.
\end{tablenotes}
\end{threeparttable}
\end{center}
\end{table*} 

\begin{table*}[htb]
\begin{center}
\begin{threeparttable}[b]
\caption{Synthesizing results for different decoders when $N=8192$ and r=0.5.}
\label{tab:Syn_result_8K}
\begin{tabular}{|c||c||c|c|c||c|c|c|}
\hline
Decoder & AE-SCL & MODE-4 & MODE-2 & MODE-1 & \cite{reduced_latency_VLSI}
& \cite{SDSCL}\tnote{$\ddagger$} & \cite{7114328}\tnote{$\dagger$}  \\ \hline
List Size & 4 & 4 & 2 & 1 & \multicolumn{3}{c|}{4} \\ \hline
Area ($\text{mm}^2$) & 4.49 & \multicolumn{3}{c||}{5.51} & 6.46 & 7.32 & 12.73 \\ \hline
Clock Rate (MHz) & 398 & \multicolumn{3}{c||}{398} & 398 & 434 & 794 \\ \hline
\# of Decoding Cycles & 2542 & 2542 & 2323 & 1975 & 2367 & 11700 & 20736\\ \hline
Latency (us) & \LTP{2542/398} & \LTP{2542/398} & \LTP{2323/398}& \LTP{1975/398}&
\LTP{2367/398} & \LTP{1170/43.4}& \LTP{2073.6/79.4}\\ \hline
NIT (Mbps) & 670 & 670 & 1473 & 3503 & 723 & 150 & 156 \\ \hline
Area Eff. (Mbps/$\text{mm}^2$) & \AEP{670/4.49} & \AEP{670/5.51} & \AEP{1473/5.51} & \AEP{3503/5.51} & \AEP{723/6.46} & \AEP{150/7.32} & \AEP{156/12.73} \\ \hline
\end{tabular}
\begin{tablenotes}
\footnotesize
\item [$\ddagger$] The decoder architecture in \cite{SDSCL} has been
  re-synthesized with the TSMC 90nm CMOS technology.
\item [$\dagger$] These results for the decoder in \cite{7114328} are
  estimated conservatively.
\end{tablenotes}
\end{threeparttable}
\end{center}
\end{table*} 

\begin{table*}[htb]
\begin{center}
\begin{threeparttable}[b]
\caption{Synthesizing results for different decoders when N=32768 and r=0.9.}
\label{tab:Syn_result_32K}
\begin{tabular}{|c||c||c|c|c|c|c||c|c|c|}
\hline
Decoder & AE-SCL & MODE-4 & \multicolumn{2}{c|}{MODE-4\_1} & MODE-2 & MODE-1 & \cite{reduced_latency_VLSI}
& \cite{SDSCL}\tnote{$\ddagger$} & \cite{7114328}\tnote{$\dagger$}  \\ \hline
List Size & 4 & 4 & 4\tnote{$\divideontimes$} & 4\tnote{$\star$} & 2 & 1 & \multicolumn{3}{c|}{4} \\ \hline
Area ($\text{mm}^2$) & 9.97 & \multicolumn{5}{c||}{11.93} & 11.89 & 15.8 & 50.41\\ \hline
Clock Rate (MHz) & 350 & \multicolumn{5}{c||}{350} & 359 & 389 & 794 \\ \hline
\# of Decoding Cycles & 6718 & 6718 & 6530 & 6206 & 6300 & 5368 & 6492 & 65813 & 96576\\ \hline
Latency (us) & \LTP{6718/350}& \LTP{6718/350} & \LTP{6530/350} & \LTP{6206/350} & \LTP{6300/350}& \LTP{5368/350}& \LTP{6492/359}& \LTP{6581.3/38.9}& \LTP{9657.6/79.4} \\ \hline
NIT (Mbps) & 1662 & 1662 & 1714 & 1811 & 3564 & 8499 & 1772 & 165 & 242\\ \hline
Area Eff. (Mbps/$\text{mm}^2$) & \AEP{1662/9.97}& \AEP{1662/11.93}&
\AEP{1714/11.93} & \AEP{1811/11.93} & \AEP{3564/11.93} & \AEP{8499/11.93} & \AEP{1772/11.89} & \AEP{165/15.8} & \AEP{242/50.41} \\ \hline
\end{tabular}
\begin{tablenotes}
\footnotesize
\item [$\ddagger$] The decoder architecture in \cite{SDSCL} has been
  re-synthesized with the TSMC 90nm CMOS technology.
\item [$\dagger$] These results for the decoder in \cite{7114328} are
  estimated conservatively.
\item [$\divideontimes$] The switching point for variation of modes is 21000.
\item [$\star$] The switching point for variation of modes is 10000.
\end{tablenotes}
\end{threeparttable}
\end{center}
\end{table*} 

Synthesis results of MM-SCL decoders for different polar codes are listed in
Tables~\ref{tab:Syn_result_1K}, \ref{tab:Syn_result_8K} and
\ref{tab:Syn_result_32K}. The decoding latency of the MODE-2 mode is
smaller than that of the MODE-4 mode and the decoder has the smallest decoding
latency with the MODE-1 mode. This is because MLD\_S1\_$q2$ and
MLD\_S1\_$q1$ have shorter data paths. Therefore, in block MM-LC-AML, three stages and
two stages of pipeline registers are used by the circuitry for the MODE-2 mode
and the MODE-1 mode, respectively. If MODE-4\_1 is used, the MM-SCL
decoder can achieve a smaller latency than the MODE-4 mode and the MODE-2 mode.

Compared with the AE-SCL decoder, the MM-SCL decoder in the MODE-4 mode has a
smaller area efficiency due to the additional circuitry for supporting multiple
modes. However, the MM-SCL decoder is more flexible to provide multiple choices
of output throughput and decoding latency, which is more suitable for variable
communication channels and applications. If a higher throughput or a smaller
decoding latency is required, the MM-SCL decoder can be
switched to the MODE-2, MODE-1 or MODE-4\_1 mode. 

Compared with the decoder in \cite{reduced_latency_VLSI}, for $N=1024$ and
$N=8192$, the MM-SCL decoder has a smaller area and a better area
efficiency. For $N=32768$, the area of the MM-SCL decoder is bigger
than that of the decoder in \cite{reduced_latency_VLSI} because the additional circuitry to
support multiple modes is larger than the saving due to the low-complexity ML
decoding unit in the MM-SCL decoder. For the (1024, 512) code, under the MODE-4,
MODE-2, and MODE-1 modes, the MM-SCL decoder provides area-efficiencies of 1.59, 3.51, and 8.39 times of
area-efficiency of the decoder in \cite{reduced_latency_VLSI}, respectively. 

Compared with decoders in \cite{SDSCL, 7114328, ParSC2}, the advantage in area
efficiency of the MM-SCL decoder is more significant. This advantage comes from two
aspects. The first is that the tree-based low-latency SCL architecture in
\cite{reduced_latency_VLSI} is adopted for the MM-SCL decoder. This helps to
reduce the decoding latency. The second is due to the low-complexity AML
decoding unit.
For $N=1024$, the MM-SCL
decoder in MODE-4 mode provides an area efficiency of 2.72, 6.77, and 2.60
times of area efficiencies of SCL decoders in \cite{SDSCL, 7114328,
  ParSC2}, respectively. When the mode is MODE-1, the ratios of the area efficiency of the
MM-SCL decoder over those of SCL decoders in \cite{SDSCL, 7114328, ParSC2} are
14.37, 35.71, and 13.73, respectively.

For $N=32768$, decoding latencies and throughputs respect to different switching points of
MODE-4\_1 are also provided. A smaller switching point leads to a smaller
latency. When the switching point is 10000, the latency of MODE-4\_1 is even
smaller than that of MODE-2. Compared with MODE-4, improvements on throughput
and latency are about 8\%.

\section{Conclusion}
\label{sec:conclusion}

In this paper, the divide-and-conquer method is applied to SC-based algorithms
in the probability domain. By extending this idea, a divide-and-conquer AML decoding unit for SCL-based polar decoder is proposed. By examining frozen-location patterns of polar codes, an efficient hardware design
for a simplified divide-and-conquer AML decoding unit is developed. To adapt to
different throughput and latency requirements, the MM-SCL polar decoder is proposed in
this paper. Synthesis results show that our implementations for our MM-SCL decoder and SCL decoder with
the LC-AML unit achieve better area efficiencies than existing SCL polar decoders.

%

\appendix
\begin{IEEEproof}[Proof of Proposition \ref{prop:1}]

\begin{enumerate}[label=(\alph*)]
\item First, $0 < z_{1,1}=2\epsilon-\epsilon^2 = 1 - (1-\epsilon)^2 < 1.$
Second, $0 < z_{1,2}=\epsilon^2 < 1$.
Then, by induction, for $i \geq 1$ and $1\leq j \leq 2^i$, $0<z_{i,j}<1$ is satisfied.
\item For any $i \geq 1$ and $1\leq j \leq 2^i$, $z_{i,2j-1} - z_{i,2j} =
2z_{i-1,j}-z_{i-1,j}^2 - z_{i-1,j}^2 = 2z_{i-1,j}(1-z_{i-1,j}).$ By Proposition~\ref{prop:1}(a), $z_{i,2j-1} - z_{i,2j} > 0 \Rightarrow z_{i,2j-1} > z_{i,2j}$.
\item By Proposition~\ref{prop:1}(b), $z_{i,4j-3} > z_{i,4j-2}$ and $z_{i,4j-1} >
z_{i,4j}$. $z_{i,4j-2} - z_{i,4j-1} =2z_{i-2,j}^2(1-z_{i-2,j})^2$. By Proposition~\ref{prop:1}(a),
$z_{i,4j-2} - z_{i,4j-1} >0 \Rightarrow z_{i,4j-2}  > z_{i,4j-1}.$ 

Therefore, $z_{i,4j-3} > z_{i,4j-2} > z_{i,4j-1} > z_{i,4j}$.

\item By Proposition~\ref{prop:1}(c), $z_{i,8j-7} > z_{i,8j-6} > z_{i,8j-5} > z_{i,8j-4}$
and $z_{i,8j-3} > z_{i,8j-2} > z_{i,8j-1} > z_{i,8j}$. We also have $z_{i,8j-5}
> z_{i,8j-3}$ and $z_{i,8j-4} > z_{i,8j-2}$ because $z_{i,4j-2} > z_{i,4j-1}$.

Now let us compare $z_{i,8j-4}$ and $z_{i,8j-3}$,
\begin{equation*}
\begin{split}
z_{i,8j-4}-&z_{i,8j-3} = -2z_{i-3,j}^2(1-z_{i-3,j})^2\\
&\times (2+4z_{i-3,j}-5z_{i-3,j}^2+2z_{i-3,j}^3-z_{i-3,j}^4).
\end{split}
\end{equation*}
By Proposition~\ref{prop:1}(a), $z_{i,8j-4} < z_{i,8j-3}$.

Therefore, $z_{i,8j-7} > z_{i,8j-6} > z_{i,8j-5} > z_{i,8j-3} > z_{i,8j-4} >
z_{i,8j-2} > z_{i,8j-1} > z_{i,8j}$.
\end{enumerate}
\end{IEEEproof}



\bibliographystyle{IEEEtran}
\bibliography{../latex/bibtex/IEEEfull,../latex/Polar}

\end{document}